\documentclass[12pt]{article}

\usepackage[dvips]{graphicx}
\renewcommand{\baselinestretch}{1.75}
\font\scaps=cmcsc10 scaled\magstep1 

\font\bigtensl=cmsl10 scaled\magstep2
\newcommand \be{\begin{equation}}
\newcommand \ee{\end{equation}}
\newcommand \ba{\begin{eqnarray}}
\newcommand \ea{\end{eqnarray}}
\begin{document}

\def\today{\ifcase\month\or
 January\or February\or March\or April\or May\or June\or
 July\or August\or September\or October\or November\or December\fi
 \space\number\day, \number\year}
%

\hfil PostScript file created: \today{}; \ time \the\time \ minutes
\vskip .15in

\centerline {DOUBLE-COUPLE EARTHQUAKE SOURCE: SYMMETRY AND
ROTATION
}

\vskip .15in
\begin{center}
{Yan Y. Kagan }
\end{center}
\centerline {Department of Earth and Space Sciences,
University of California,}
\centerline {Los Angeles, California 90095-1567, USA;}
\centerline {Emails: {\tt ykagan@ucla.edu,
kagan@moho.ess.ucla.edu }}
\vskip 0.02 truein

\vspace{0.15in}

\noindent
{\bf Abstract.}
We consider statistical analysis of double couple ($DC$)
earthquake focal mechanism orientation.
The symmetry of $DC$ changes with its geometrical properties,
and the number of 3-D rotations one $DC$ source can be
transformed into another depends on its symmetry.
Four rotations exist in a general case of $DC$ with the
nodal-plane ambiguity, two transformations if the fault plane
is known, and one rotation if the sides of the fault plane are
known.
The symmetry of rotated objects is extensively analyzed in
statistical material texture studies, and we apply their
results to analyzing $DC$ orientation.
We consider theoretical probability distributions which can be
used to approximate observational patterns of focal
mechanisms.
Uniform random rotation distributions for various $DC$ sources
are discussed, as well as two non-uniform distributions: the
rotational Cauchy and von Mises-Fisher.
We discuss how parameters of these rotations can be estimated
by a statistical analysis of earthquake source properties in
global seismicity.
We also show how earthquake focal mechanism orientations can
be displayed on the Rodrigues vector space.
\vskip .15in
\noindent
{\bf Short running title}:
{\sc
Double-couple: Symmetry and rotation
}

\vskip .25in

\section{Introduction}
\label{intro} This paper addresses two problems: the random rotation
of double-couple ($DC$) earthquake sources and how symmetry
properties of these sources influence their rotation angle
distribution and their display. Properties of earthquake focal
mechanisms and methods for their determination are considered by
Snoke (2003) and Gasperini \& Vannucci (2003). Ekstr\"om {\it et
al.}\ (2005, and references therein) discuss their extensive work on
evaluating seismic moment tensors for global earthquakes.

In this paper we considered only the double-couple earthquake
focal mechanism.
For tectonic events non-double-couple mechanisms like the CLVD
are likely due to various systematic and random errors in
determining the mechanism (Frohlich \& Davis 1999; Kagan 2003,
2009).
These results suggest that routinely determined CLVD values
would not reliably show the deviation of earthquake focal
mechanisms from a standard $DC$ model.

Snoke (2003) and Gasperini \& Vannucci (2003) consider several
equivalent representations for double-couple sources and their
properties, and provide mathematical expressions for their
mutual transformation.
Krieger and Heimann (2012, and references therein) review
routines for plotting moment tensors and focal mechanisms.

Two general techniques can be employed to study the 3-D
rotation: orthonormal rotation matrices and normalized
(unit) quaternions.
The quaternion method has been used to evaluate these
rotations in many investigations of earthquake focal
mechanisms (see, for example, Kagan 1991; Frohlich \& Davis
1999; Kagan 2009; Kagan \& Jackson 2011).
Kagan (2007) explains how `ordinary' matrices and vectors
can be used to obtain 3-D rotation parameters.
Below we comment on the advantages and drawbacks of both
methods.

Altmann's (1986) book was a first monograph specifically
dedicated to 3-D rotations [group $SO(3)$] and quaternions.
At present quaternions are widely used to describe rotations in
space satellite and airplane dynamics (Kuipers 1999) and
simulations of virtual reality, robotics and automation
(Hanson 2006; Dunn \& Parberry 2011).
These last three monographs explain quaternions in a more
accessible manner.
Many journal articles (see references in these monographs and
in Kagan 2009) discuss practical application of quaternions
for analyzing the 3-D rotations.

However, the above publications do not consider the symmetry
properties of rotated objects or how symmetry influences orientation
analysis. As Kagan (1990; 1991; 2007; 2009) indicated, the
techniques considered in those publications cannot be used for the
$DC$ source orientation studies without major modifications, because
of the $DC$ symmetry properties, described in Section~\ref{symm}.
The only scientific discipline where symmetry is extensively
considered in 3-D rotation analysis is study of material texture
(Handscomb 1958; Mackenzie 1958; Grimmer 1979; Frank 1988; Heinz \&
Neumann 1991; Morawiec, 2004; Meister \& Schaeben 2005; Schaeben
2010). In this paper (Section~\ref{text}) we discuss applying these
results to $DC$ source investigations. In Section~\ref{dcr1} we
consider theoretical probability distributions used to approximate
observational patterns of focal mechanisms. Section~\ref{stat} is
dedicated to the statistical analysis of earthquake source
properties in global seismicity. Using the results of
Section~\ref{text} we also show how earthquake focal mechanism
orientations can be displayed in the Rodrigues vector space.
Section~\ref{disc} and Section~\ref{conc} summarize our results.

\section{Focal mechanism symmetry }
\label{symm} Depending on the known properties of double-couple
earthquake focal mechanism, we consider three types of earthquake
source symmetry (Kagan 1990): \hfil\break $\bullet$ 1. $DC1$ --
double couple with no symmetry or the identity ($I$) symmetry, if
the focal plane and its sides are known; \hfil\break $\bullet$ 2.
$DC2$ -- double couple with $C_2$, order 2 cyclic symmetry, i.e.,
the focal plane is known, but its sides are not; \hfil\break
$\bullet$ 3. $DC4$ -- double couple with nodal-planes that not
distinguishable; it has $D_2$, order 2 dihedral symmetry.

These earthquake source symmetries correspond to the following
crystallographic symmetries considered in material texture
analysis (see, for instance, Morawiec 2004).
$DC4$ has an orthorhombic symmetry (as in a rectangular right
parallelepiped or a rectangular box with unequal sides);
$DC2$ has a monocline symmetry (as in a 3-D prism with two
angles of 90$^\circ$ and one arbitrary angle); $DC1$ has a
triclinic, or no symmetry.

\begin{figure}
\begin{center}
\includegraphics[width=0.75\textwidth]{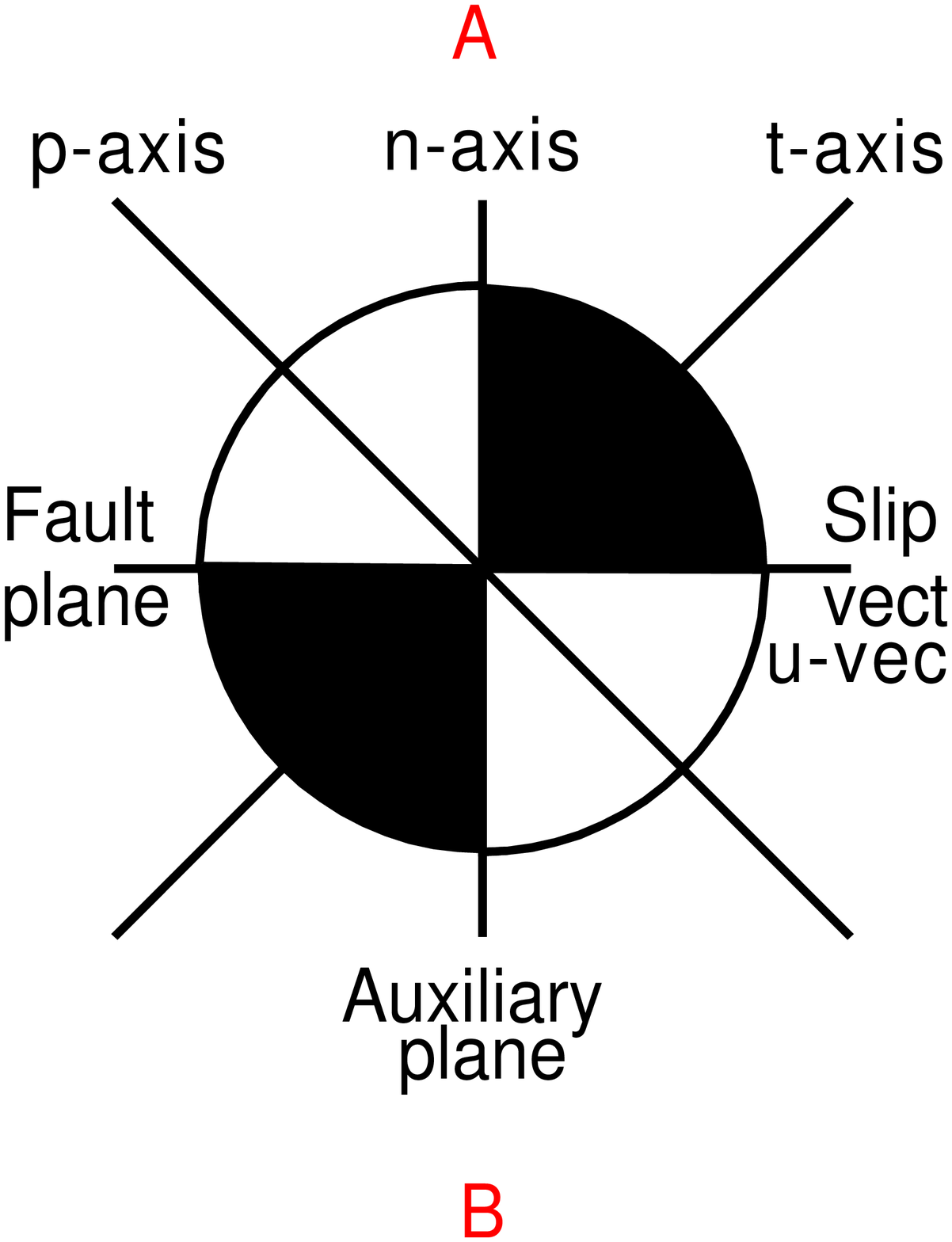}
\caption{\label{fig01}
}
\end{center}
Schematic (beachball) diagram of the $DC$ earthquake focal
mechanism and its quadrupole radiation patterns.
The null ({\bf b}) axis is orthogonal to the {\bf t}- and {\bf
p}-axes, or it is located on the intersection of fault and
auxiliary planes, i.e., perpendicular to the paper sheet in
this display.
The {\bf n}-axis is normal to the fault-plane; {\bf u} is
a slip vector.
\end{figure}

Fig.~\ref{fig01} displays the geometry of the $DC$ source (Aki
\& Richards 2002).
It represents the quadrupolar `beachball' radiation patterns
of earthquakes.
The focal plots involve painting on a sphere the sense of the
first motion of the primary P-waves: solid for compressional
motion and open for dilatational.
The two orthogonal nodal planes separating these areas are
the fault and the auxiliary planes.
During routine determination of focal mechanisms, it is
impossible to distinguish between these planes, a property
called `nodal-plane ambiguity.'
The planes' intersection is the null-axis (called ${\bf
b}$-axis), the ${\bf p}$-axis is in the middle of the open
lune, and the ${\bf t}$-axis is in the middle of the closed
lune.
These three axes are called the `principal axes of an
earthquake focal mechanism,' and their orientation defines the
mechanism.

To make the focal mechanism picture unique, the eigenvectors
are pointed down in regular representations.
However, the handedness of the coordinate system formed by the
vectors can change as the result of such an assignment.
The systems of the opposing handedness cannot be rotated one
into another.
In most of our considerations, we use the right-handed
coordinate system placed at each earthquake centroid.

\begin{figure}
\begin{center}
\includegraphics[width=0.75\textwidth]{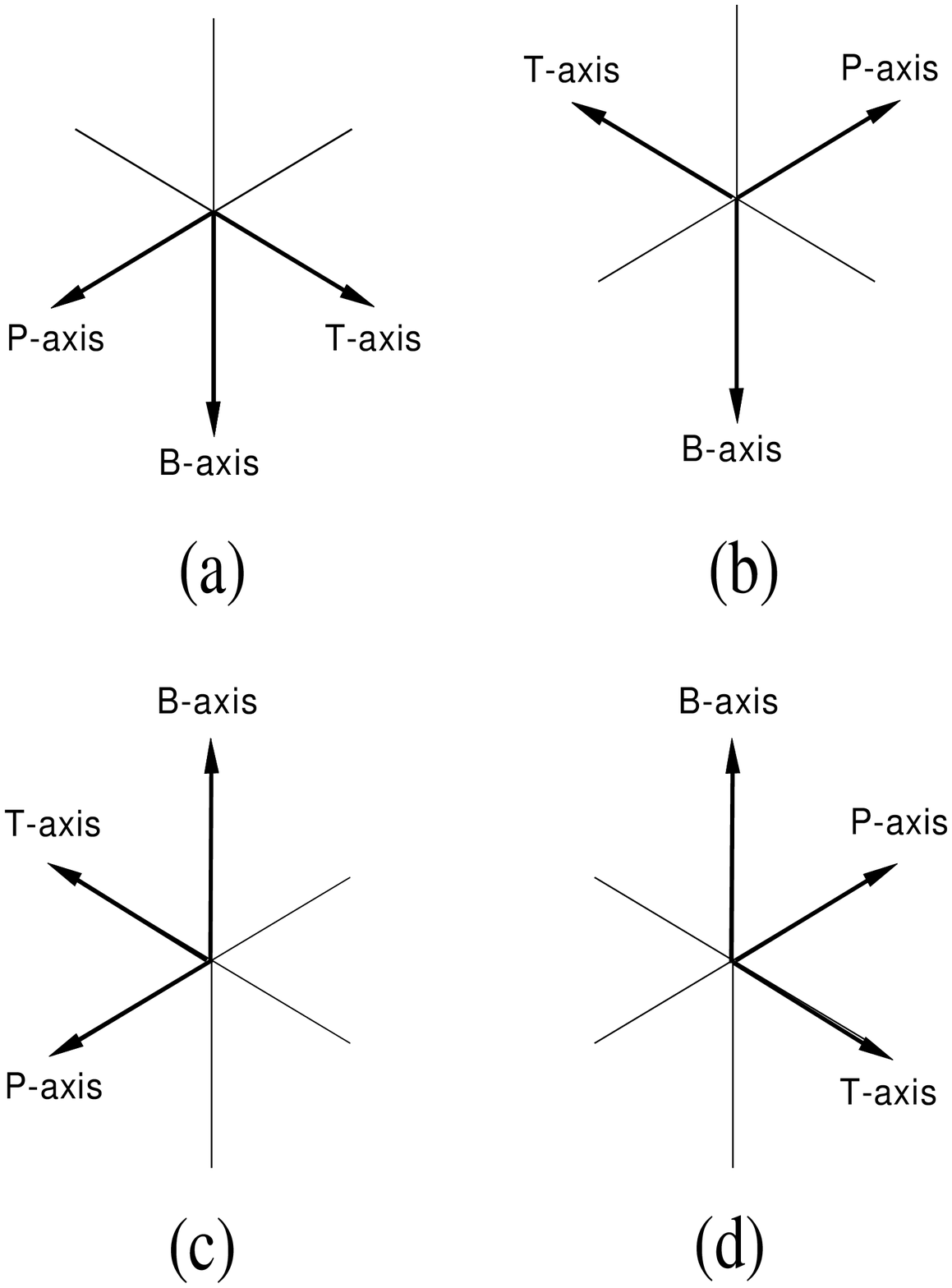}
\caption{\label{fig02}
}
\end{center}
Four schematic diagrams of earthquake focal mechanism, having
the $DC4$ symmetry.
The right-hand coordinate system is used.
\end{figure}

Fig.~\ref{fig02} displays four examples of the
right-handed coordinate system for a $DC4$ source.
The system can be arbitrarily rotated, and the handedness of
the system is preserved.
The left-handed system can be obtained in this picture if one
inverts the direction of any individual axis or of all three
axes.
If the direction of two axes is reversed, the handedness of
the system is preserved.

The earthquake focal plane can often be determined by
inverting the higher-rank point seismic moment tensors
(McGuire {\it et al.} 2001; Chen {\it et al.} 2010) or by the
aftershock pattern.
The face/side (up/down or foot/hanging wall) of a
focal plane generally is unknown.
In such a case the $DC$ focal mechanism has a $C_2$ symmetry;
we call it $DC2$.

\begin{figure}
\begin{center}
\includegraphics[width=0.75\textwidth]{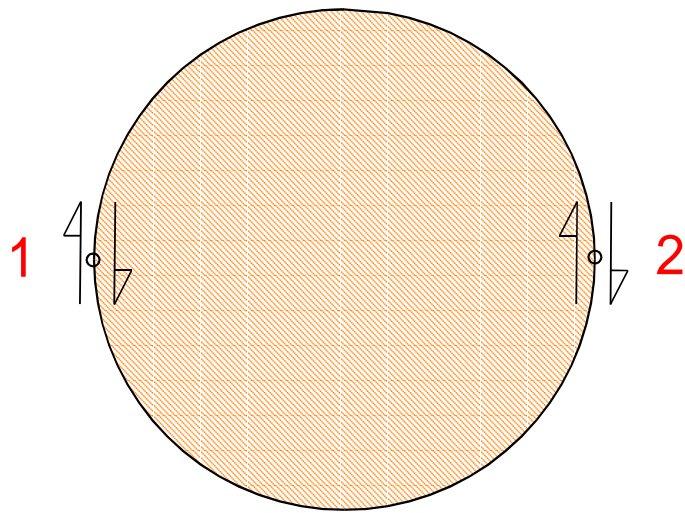}
\caption{\label{fig03}
Schematic diagram of earthquake focal mechanism $DC2$.
}
\end{center}
\end{figure}

Finally, the face or the side of the focal plane can be known;
it is shown in Fig.~\ref{fig01} by symbols $A$ and $B$.
Such a source is called $DC1$.
Fig.~\ref{fig03} illustrates the difference between $DC2$ and
$DC1$ mechanisms.
If a cylinder of one material is rotated in a half-space of
another material, the two $DC1$ sources `1' and `2' would have
a rotation angle of 180$^\circ$.
If they are considered as $DC2$ sources, two angles 0$^\circ$
and 180$^\circ$ are possible to rotate one mechanism into
another.

In Fig.~\ref{fig02}, any of the configurations appropriately
rotated can represent a $DC1$ source, but only pairs (a)-(b)
or (c)-(d) correspond to a $DC2$ mechanism.
All four diagrams correspond to a $DC4$ source.

\section{Earthquake focal mechanism and material texture
statistics }
\label{text}
Frank (1988) proposes using the Rodrigues vector space to
represent 3-D rotation of symmetrical objects.
This representation has an advantage: under any transformation
of the Rodrigues map corresponding to a change of the
reference orientation, straight lines transform into straight
lines, and planes into planes.

For an object with the non-identity symmetry, accepted points
lie in a region around the origin, which is called `the
fundamental zone of the map.'
It is a polyhedron, bounded by the planes which are
orientationally equidistant between the origin and the
neighboring equivalent point by a symmetry rotation to the
origin.
Any points lying outside one of these planes have an
equivalent point lying inside the fundamental zone (Frank
1988).
For an orthorhombic crystal with three orthogonal axes, the
fundamental zone is a cube, with its six faces orthogonal to
the axes at a distance from the origin of $\tan \pi/4 = 1$
(Frank 1988).
The cube is surrounded by three neighboring zones, each
divided into two at infinity.

Fig.~\ref{fig04} shows the fundamental zone for a $DC4$
source (Heinz \& Neumann, 1991, Fig.~7).
It is a cube with corner coordinates $ x_1 = \pm 1.0; \ x_2 =
\pm 1.0; \ x_3 = \pm 1.0$.
Owing to the $DC4$ symmetry, an octant of the cube contains
full information about the orientation distribution for
uniformly random rotation.
This octant is called the `MacKenzie cell' (Morawiec \& Field
1996, see also their Fig.~1 displaying the cells for the $D_3$
and $D_4$ symmetries).

Each point {\sl inside} the cube uniquely corresponds to a
certain orientation/rotation with a minimum rotation angle
$\Phi_{\rm min} \le 120^\circ$.
The points in the inscribed sphere of the cube correspond to
the rotations with angles $\Phi_{\rm min} \le 90^\circ$.
The other three rotation angles are situated outside the
fundamental zone.
For example, the point of the zero rotation is located at the
cube center, whereas three other rotation points are at
infinity: $x_1 = \pm \infty; x_2 = \pm \infty; x_3 = \pm
\infty$, corresponding to 180$^\circ$ rotations.
These points at $\pm \infty$ are equivalent (Altmann 1986).
Similarly, for any point inside the cube, three points outside
correspond to the rotations with angles $\Phi > \Phi_{\rm
min}$.

However, when a point moving orthogonally from the origin
reaches a cube face, it simultaneously appears on the opposite
face: two 90$^\circ$ rotations produce the same effect
(Fig.~\ref{fig05}).
This means that when we determine the minimum angle
$\Phi_{\rm min}$ for cyan point rotations shown in
Fig.~\ref{fig05} using the program developed by Kagan (1991),
we find two equal solutions.
The remaining two angles are greater than $\Phi_{\rm min}$
(see also the next Section).

If a point on one face moves to an edge, the `identical' point
on the opposite face simultaneously moves to another edge
until both points reach the middle of the edges.
This orientation corresponds to the rotation $\Phi \approx
109.5^\circ$, and Fig.~\ref{fig06} shows that there are three
equivalent points at the edges.
The third point appears as it moves from the outside of the
cube to the third edge.
As in Fig.~\ref{fig05}, this means that three equal
angles $\Phi_{\rm min}$ would be obtained.
Finally, when a point is at a vertex, as shown in
Fig.~\ref{fig07}, three other vertices correspond to the same
rotation $\Phi = 120^\circ$ (Frank 1988), i.e., all four
rotation angles are $\Phi_{\rm min}$.

This arrangement of the orientations for the rotation angles
$\Phi \ge 90^\circ$ describes a complex topology for $DC4$
source rotation.
This topology involves projective or M\"obius transformation
(Altmann 1986; Frank 1988).
Full analysis of the $DC4$ source orientation, when and if
performed, would involve very intricate investigations of
rotation angle transformations due to source symmetry.

The Rodrigues space has no special advantages in displaying
the orientation distribution for two other sources: $DC2$ and
$DC1$.
For $DC2$ the fundamental zone overlaps the entire Rodrigues
space.
For $C_2$ symmetry the fundamental zone is bounded by two
planes perpendicular to the {\bf b}-axis, each at the distance
$\tan(\pi/4) = 1$ from 0.
For a $DC1$ source, the whole Rodrigues space up to infinity
is included.
In these cases other spaces are more convenient in displaying
the 3-D rotation distribution (Frank 1988; Morawiec \& Field,
1996).
Altmann (1986, pp.~164-176) explains the projective or
M\"obius topology of rotations in the quaternion parametric
ball for non-symmetrical objects.

\section{Rotation of double couple earthquake sources }
\label{dcr1}
\subsection{$DC$ symmetry and rotation angle }
\label{dcr1a}
In our earlier paper (Kagan 1991) we considered the inverse
problem of the $DC4$ source rotation, i.e., given two
earthquake focal mechanisms to determine all the 3-D rotations
by which one mechanism can be rotated into another.
As we showed, the $DC4$ symmetry results in four
such rotations with angle $\Phi$ range $0^\circ \le \Phi <
180^\circ $ ({\sl cf.} Fig.~\ref{fig02}).
For most practical purposes, the rotation with a minimum angle
$\Phi_{\rm min}$ can be selected.
Following the material texture designation (Morawiec 2004,
p.~115), we sometimes refer to general rotation angles as
`misorientation angles' and the minimum rotation angle as the
`disorientation angle.'

Here we consider how to compute the rotation of a $DC1$ source
needed to align it with some reference $DC1$ source.
It is unlikely that sufficient data would exist on $DC1$
sources for a statistical study of their distribution.
However, in some cases we need to measure their angle of
disorientation or the angular distance.
Moreover, if the disorientation angle between two $DC1$
sources is large, it would be almost impossible to identify
their fault planes and plane faces.
Small rotation angles for such focal mechanisms are likely to
correspond to $\Phi_{\rm min}$ for the $DC4$ source, as
mentioned above.

To compute the disorientation of a $DC1$ source, we can modify
our {\scaps fortran} program listed in Kagan (1991).
When a fault plane and its face are known, a focal mechanism
would be better specified through a fault plane geometry (Aki
\& Richards 2002, Figs.~4.13 and 4.20) with three angles:
strike or azimuth ($\phi$), dip ($\delta$), and rake
($\lambda$).
Usually the range of these angles is taken as follows:
$0^\circ \le \phi < 360^\circ $, $0^\circ \le \delta <
90^\circ $, $-180^\circ \le \lambda < 180^\circ $.
The problem arises when comparing two sources if the dip
($\delta$) of one focal plane exceeds 90$^\circ$, so that a
foot wall of one mechanism becomes a hanging wall for another
source (Aki \& Richards 2002).
To simplify the calculations in our program (see below), we
extend the $\delta$ range to $180^\circ$.

If the face/side of a fault plane is unknown, as shown in
Fig.~\ref{fig03}, we need to calculate the second angle of
the rotation for the $DC2$ source.
As with the $DC1$ source, the data on $DC2$ sources are sparse
and insufficient for a statistical study, but we need a
technique to measure their angles of disorientation.
An easy way to accomplish this measurement would be to change
the strike of the fault plane by $180^\circ$ and change the
rake sign. The modified {\scaps dc1rot.for} program is
available at
http://jumpy.igpp.ucla.edu/$\sim$kagan/dc1rot.for.

In this program we use the quaternion technique to determine
the rotation angle and the rotation axis parameters to
transform one $DC$ source into another.
Quaternions are used because for rotation angle $\Phi$ close
or equal to $180^\circ$, the matrix method cannot determine
the rotation axis parameters (Kagan, 2007).

Kagan (2009, Appendices $A$ and $B$) discusses normalized
quaternions and their relation to the $DC4$ source.
Representations for the $DC4$ by seismic moment tensors as
well as by orthonormal matrices are considered.
A quaternion representation allows a relatively easy
determination of the rotation angle and the axis parameters
for the $DC4$ earthquake focal mechanism (Kagan 1991).

If only the rotation angle is needed, then one can use a
scalar (dot) product of two quaternions (Hanson 2006, p.~65;
Dunn \& Parberry 2011, p.~255) to determine the angle:
\be
\cos(\Phi/2) \ = \ {\bf q}^a \cdot {\bf q}^b \ = \
q^a_1 \, q^b_1 +
q^a_2 \, q^b_2 +
q^a_3 \, q^b_3 +
q^a_4 \, q^b_4
\, ,
\label{eq0}
\ee
where ${\bf q}^i$ are normalized quaternions for each $DC$
source and $q_j$ are the quaternion's components.

In our program we first compute the orthonormal matrix for
each $DC$ source and then determine the corresponding normalized
quaternion.
There is a possibility of losing precision when converting a
matrix to a quaternion (Shepperd 1978; Horn 1987).
A certain computation technique should be applied to avoid
this. We used a similar technique in our programs {\scaps
dcrot.for} (Kagan 1991, the end of the {\scaps subroutine
quatfps}) and in {\scaps dc1rot.for} (see above).

\subsection{Rotation angle distribution }
\label{dcr1b}

\subsubsection{Uniform random rotation of $DC$ sources }
\label{dcr1b1}

The distribution of the uniform random rotation for $DC$
sources constitutes a reference for distributions occurring in
earthquake focal mechanisms where we expect the distributions
to be partially random.
These stochastic distributions can be analytically calculated
by taking into account the sources' symmetry.

A distribution of the minimum angle $\Phi_{\rm min} $ for a
uniform random rotation of the $DC4$ source was obtained by
Kagan (1990, Eqs.~3.1-3.3), using the results by Handscomb
(1958) and Mackenzie (1958) for the random disorientation of
two cubes.
The probability density function (PDF) is
\be
f(\Phi) \ = \ (4/\pi)(1 - \cos \Phi) \quad {\rm for} \quad
0 \le \Phi \le \pi/2
\, ;
\label{eq1}
\ee
\be
f(\Phi) \ = \ (4/\pi) (3 \sin \Phi + 2 \cos \Phi - 2)
\quad {\rm for} \quad
\pi/2 \le \Phi \le \Phi_S
\, ;
\label{eq2}
\ee
and
\ba
f(\Phi) \ = \ & (4/\pi) \, \Bigg \{ 3 \sin \Phi + 2 \cos \Phi
- 2 \, -
\nonumber \\
& (6/\pi) \bigg [ 2 \, \sin \Phi \, \arccos \left ({{1 + \cos
\Phi} \over {- 2 \cos \Phi}} \right )^{1/2} -
\nonumber \\
& ( 1 - \cos \Phi) \, \arccos {{1 + \cos \Phi}
\over {-2 \cos \Phi}} \, \bigg] \Bigg \}
\nonumber \\
& \quad {\rm for} \quad
\Phi_S \le \Phi \le {{2\pi} \over 3} \, ,
\label{eq3}
\ea
where
\be
\Phi_S \ = \ 2 \arccos \, (3^{-1/2})
\ = \ \arccos \, \left (- {1 \over 3} \right ) \ \approx \
109.47^\circ \, .
\label{eq4}
\ee
For the $DC2$ source a similar PDF is
\be
f(\Phi) = (2/\pi) [1 - \cos(\Phi)]  \quad {\rm for} \quad 0
\le \Phi \le \pi/2
\, ;
\label{eq5}
\ee
and
\be
f(\Phi) = (2/\pi) \sin(\Phi)  \quad {\rm for} \quad \pi/2
\le \Phi \le \pi \, .
\label{eq6}
\ee
For the $DC1$ source the function is
\be
f(\Phi) = (1/\pi) [1 - \cos(\Phi)]  \quad {\rm for} \quad 0
\le \Phi \le \pi \, .
\label{eq7}
\ee

Grimmer (1979), also following Handscomb (1958) and Mackenzie
(1958) results, obtained similar analytic expressions for a
completely random rotation of orthorhombic, monoclinic, and
triclinic crystals (equivalent in symmetry to the $DC$
earthquake source with various restrictions described above).
He listed median angles as well as their mean and standard
deviations for all these distributions.
Morawiec (1995; 2004, pp.~117-119) derived these distributions
by integration in the Rodrigues space.

\subsubsection{Non-uniform distributions of random rotations }
\label{dcr1b2}

Two non-uniform rotation angle distributions are useful in
analyzing earthquake focal mechanism rotation: the rotational
Cauchy law (Kagan 1982, 1992) and von Mises-Fisher/Bingham
rotational distribution (Kagan 1992, 2000; Schaeben 1996;
Mardia \& Jupp 2000, pp.~289-292; Morawiec 2004, pp.~88-89).

The Cauchy distribution is especially important for
representing earthquake geometry, since it can be shown by
theoretical arguments (Zolotarev 1986, pp.\ 45-46; Kagan,
1990) and simulations (Kagan 1990) that the stress tensor in a
medium with random defects follows this distribution.
The Cauchy law is a {\it stable} distribution (Zolotarev,
1986).
The stable distributions are essential for two reasons:
a) They are invariant under addition of random variables;
b) Stable distributions have a power-law tail, i.e., they
are asymptotically scale-invariant.

The probability density function (PDF) of the rotational
Cauchy distribution can be written as (Kagan 1982; 1990)
\be
f(\Phi) \ = \ {{2} \over {\pi}} \Biggr [
{{\kappa A^2 (1 + A^2) } \over
{(\kappa^2 + A^2)^2}} \Biggl ]
\ = \
{{4\,\kappa\,\left[ 1 - \cos ({\Phi}) \right]
}\over {\pi \,{{\left[ 1 + {\kappa^2} + \left(
{\kappa^2} - 1 \right) \,\cos ({\Phi}) \right] }^2}}} \, ,
\quad {\rm for} \quad
180^\circ \ge \Phi \ge 0^\circ \, ,
\label{eq8}
\ee
where $A = \tan(\Phi/2)$.
The scale parameter $\kappa$ of the Cauchy distribution
represents the degree of {\it incoherence} or {\it complexity}
in a set of earthquake focal mechanisms.
The cumulative rotational Cauchy distribution can be written
as
\be
F(\Phi) = {{2} \over {\pi}} \Biggr [ \arctan(A/\kappa) -
{{A\times \kappa} \over {A^2 + \kappa^2}} \Biggl ] \, .
\label{eq9}
\ee

The Cauchy distribution is assumed to be axisymmetric on the
quaternion hypersphere $S^3$.
This means that the rotation axis poles are distributed
uniformly over a regular $S^2$ sphere.
For a general case, the axes distribution for earthquake focal
mechanisms may need to be specified as non-uniform.
In that case certain rotations would be preferred depending on
the focal mechanism of a reference event.
However, we have not yet advanced to this stage (see
Section~\ref{stat}).

The von Mises-Fisher/Bingham distribution for the 3-D
orientation is widely discussed in literature (Schaeben 1996;
Mardia \& Jupp 2000; Morawiec 2004).
Schaeben (1996) and Morawiec (2004) show that this
distribution is essentially equivalent to the Bingham
distribution.
The von Mises-Fisher distribution is a Gaussian-shaped function
concentrated near the zero angle.
This distribution can be implemented to model random errors in
determining focal mechanisms.
The distribution has many forms.
However, even the simplest axially symmetric expressions, due
to the complexity of normalization, represent difficult
computation.

For small values of the standard error $\sigma_\Phi$, the von
Mises-Fisher-type distribution is equivalent to the rotational
Maxwell law used by Kagan \& Knopoff (1985) and Kagan (1992).
The latter distribution is obtained by generating a 3-D
normally distributed random variable {\bf u} ($u_1$, $u_2$,
$u_3$) with the standard deviation $\sigma_{\bf u}$
($\sigma_{u_1}, \, \sigma_{u_2}, \, \sigma_{u_3}$) and then
calculating the unit quaternion
\ba
q_0 \ = \ 1/ \sqrt { 1 + u_1^2 + u_2^2 + u_3^2 } \, , &&
\nonumber\\
q_i \ = \ u_i / \sqrt { 1 + u_1^2 + u_2^2 + u_3^2 } \, ,
\quad {\rm for} \ i = 1, 2, 3. &&
\label{eq10}
\ea
The 3-D rotation angle is calculated
\ba
\Phi \ = \ & 2 \arccos (q_0)
\ \approx \ 2 \arccos \Bigl (1 - (u_1^2 + u_2^2 + u_3^2)/2
\Bigr )
\nonumber\\
\approx \ & 2 \arcsin \sqrt {u_1^2 + u_2^2 + u_3^2} \
\approx \ 2 \sqrt {u_1^2 + u_2^2 + u_3^2} \, .
\label{eq11}
\ea
The final expression is twice the length of a vector in the 3-D
space.

Since components of vector {\bf u} are normally
distributed, the angle $\Phi$ (in degrees) follows the Maxwell
distribution with
\be
\sigma_\Phi \ = \ 360 \, \sigma_u / \pi \, ,
\label{eq12}
\ee
where we assume that all components of $\sigma_{\bf u}$ are
equal ($\sigma_{u_1} = \sigma_{u_2} = \sigma_{u_3} =
\sigma_u$).
For $180^\circ \ge \Phi \ge 0^\circ$ the Maxwell PDF is
\be
\psi (\Phi) \ = \ \sqrt {2 \over \pi} \times
{ {\Phi^2} \over {\sigma_\Phi^3}}
\times \exp \left [-\Phi^2/(2 \sigma_\Phi^2) \right ] \, .
\label{eq13}
\ee
This equation describes the distribution of a vector length in
three dimensions, if the vector components have a Gaussian
distribution with a zero mean and a standard error
$\sigma_\Phi$.
The Maxwell cumulative distribution function (CDF) is
\be
\Psi (\Phi) \ = \ {\rm erf} \left ( { \Phi \over
{ \sigma_\Phi \sqrt 2 } } \right ) \ - \
\sqrt {2 \over \pi} \times
{ {\Phi} \over {\sigma_\Phi}}
\times \exp \left [-\Phi^2/(2 \sigma_\Phi^2) \right ] \, ,
\label{eq14}
\ee
where ${\rm erf} \, (.)$ is an error function.

The major problem with these non-uniform random distributions
is that they do not consider the symmetry of the rotated
object.
When rotation angles are small, the distribution is
concentrated around the zero angle neighborhood.
For the $DC4$ source as shown in Fig.~\ref{fig04}, almost all
distribution density would be inside the fundamental zone.
However, for more spread out angle distributions, we should
account for cases where the rotation angle exceeds the maximum
angles (see, for example, Eqs.~\ref{eq1}-\ref{eq4}).
Then the distribution would be folded back into the
fundamental zone.

Mason \& Schuh (2009) propose to convolve angle distribution with
appropriate 3-D spherical or 4-D hyperspherical harmonics to obtain
a new angle distribution which fits into the fundamental zone. It is
not clear whether such calculations can be made analytically.
Simulation seems the only practical way to transform both Cauchy and
von Mises/Fisher distributions for the $D_2$ symmetric case (i.e.,
for maximum $\Phi_{\rm min}$ rotation angle 120$^\circ$). Kagan
(1992, Fig.~3c) produced such distributions. Fig.~\ref{fig08} below
plots the appropriate Cauchy distribution which is reduced to
$\Phi_{\rm min} \le 120^\circ$.

\section{Focal mechanisms statistics }
\label{stat}
\subsection{Disorientation angle statistics }
\label{stat1}
Since there is no general model of earthquake focal mechanism
distribution, we need to study the distribution of mechanisms
in earthquake catalogs empirically to infer their properties.
How various tectonic and geometrical factors shape the
distribution of earthquake sources should be studied as well.
Such investigations are difficult because we are dealing with
a multidimensional stochastic point process: earthquake size,
occurrence time, location, and source parameters serve
as potential inputs to the distributions.

In this paper we are mostly interested in the distributions of
rotation between two earthquake focal mechanisms.
Even if we fix earthquake time, space, and magnitude interval,
the $DC$ rotation distribution depends on at least three
variables: the rotation angle and two spherical coordinates of
a rotation axis pole.
Displaying all three degrees of freedom in a distribution
presents a difficult problem.
Therefore, in our previous investigations we studied partial
distributions.
For example, Kagan (1992, Figs.~6-9; 2009, Figs.~9,10)
obtained various distributions of the rotation angle
$\Phi_{\rm min}$ between two focal mechanisms.
Below we first update our most important results on the
distribution of the rotation angle $\Phi_{\rm min}$, and then
analyze a three-dimensional distribution of rotation angle and
the axes in the Rodrigues space.

We used the Global Centroid Moment Tensor catalog (Ekstr\"om
{\it et al.} 2005), referred to subsequently as GCMT.
This catalog employs relatively consistent methods and reports
tensor focal mechanisms.
The GCMT catalog started in 1977, and is complete only
for earthquakes with magnitudes of about 5.8 and larger.
The present catalog contains more than 36,000 earthquake
entries from 1977/1/1 to 2011/12/31.

Fig.~\ref{fig08} displays cumulative distributions of the
rotation angle $\Phi_{\rm min}$ for shallow earthquake pairs
with the magnitude threshold $m_t = 5.0$ that are separated by
a distance of less than 50~km.
We study whether the rotation of focal mechanisms depends on
where the second earthquake of a pair is situated with regard
to the first event.
Thus, we measure the rotation angle for centroids located in
30$^\circ$ cones around each principal axis of the first event
(see curves, marked the {\bf t}-, {\bf p}-, and {\bf b}-axes).

The curves in Fig.~\ref{fig08} are narrowly clustered, with about
95\% of angles less than 90$^\circ$, within an inscribed sphere of
the fundamental zone (Fig.~\ref{fig04}). This pattern can be
compared to the uniform rotation (Eqs.~\ref{eq1}-\ref{eq3}) for
which 72.7\% ($2-4/\pi$) of angles are within 90$^\circ$. The curves
are obviously well approximated by the $DC$ rotational Cauchy
distribution (Eq.~\ref{eq8}). This distribution is characterized by
a parameter $\kappa$; a smaller $\kappa$-value corresponds to the
rotation angle $\Phi_{\rm min}$ concentrated closer to zero. Thus,
regardless of spatial orientation, all earthquakes have focal
mechanisms similar to a nearby event. Earthquakes in the cone around
the {\bf b}-axis correspond to a smaller $\kappa$-value than events
near the other axes. These results are similar to those shown in
Fig.~6 by Kagan (1992) or Fig.~9 by Kagan (2009).

Fig.~\ref{fig09} and Fig.~\ref{fig10} show the disorientation
angle distribution for the magnitude cutoff $m_t = 5.8$.
As may be expected for the higher magnitude, the angles are
concentrated closer to zero, and the difference between the
curves corresponding to various cones increases.
Maxwell distribution curves are shown to illustrate possible
behavior of the angle distributions near zero.
The $\sigma_\Phi = 7.5^\circ$ parameter of the distribution is
small, compared to the distribution range ($120^\circ \ge \Phi
\ge 0^\circ$).
Therefore, the curves are concentrated close to zero; we do
not need simulation to consider the curve behavior for large
values of $\Phi_{\rm min}$, as done, for example, in Fig.~3c
by Kagan (1992).

The difference in the distribution curves corresponding to
various focal mechanism axes suggests that the Cauchy
distribution parameter $\kappa$ depends upon the geometry
of a fault system.
Contrary to our assumptions (Eq.~\ref{eq8}), poles of rotation
axes are not uniformly distributed over the $S^2$ sphere.

\subsection{Distributions of rotation axes }
\label{stat2}

\begin{figure}
\begin{center}
\includegraphics[width=0.75\textwidth]{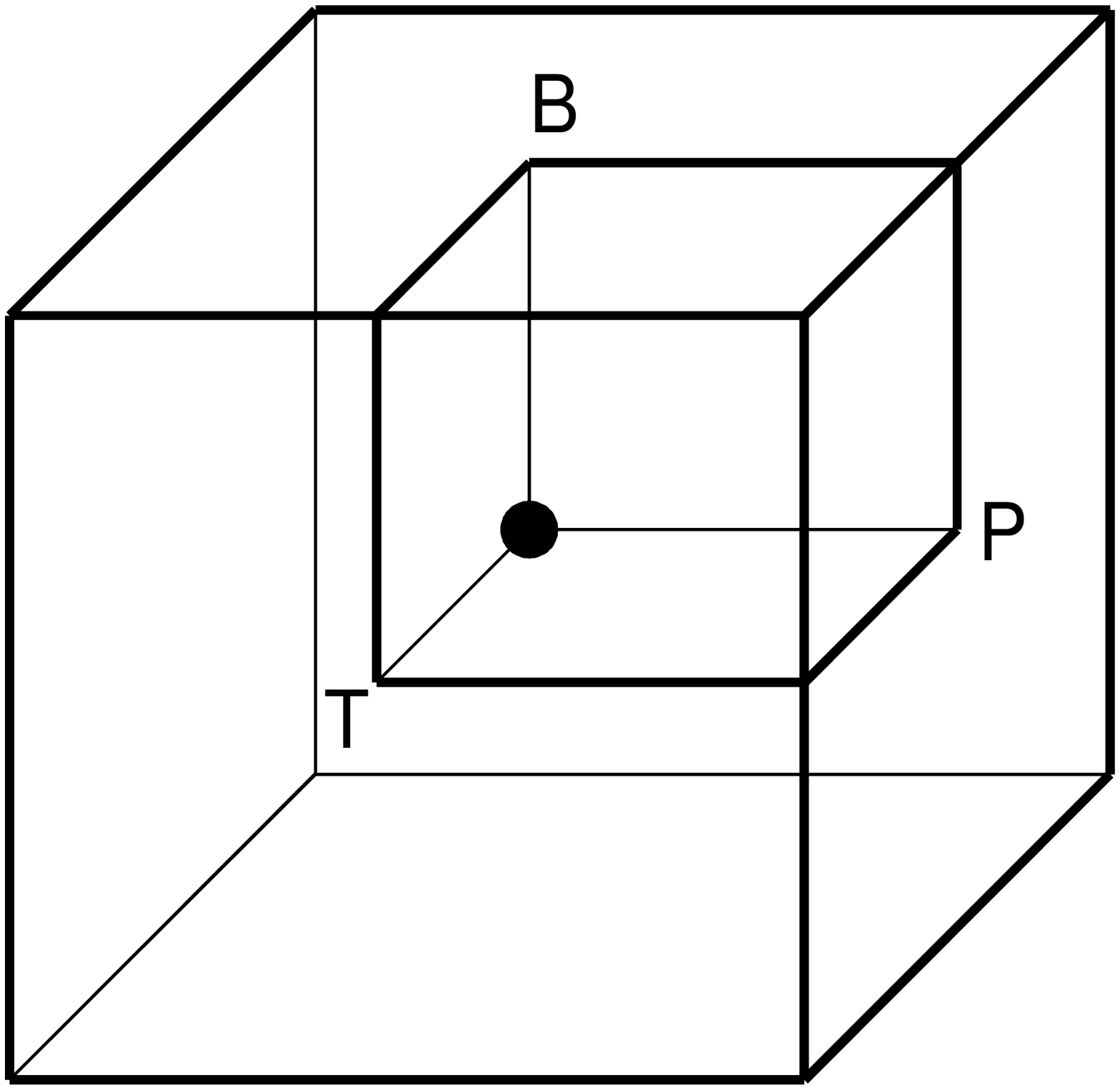}
\caption{\label{fig04}
}
\end{center}
Fundamental zone display for $DC4$ source.
The ${\bf b}$, ${\bf p}$, and ${\bf t}$-axes of the source
are shown.
\end{figure}

\begin{figure}
\begin{center}
\includegraphics[width=0.75\textwidth]{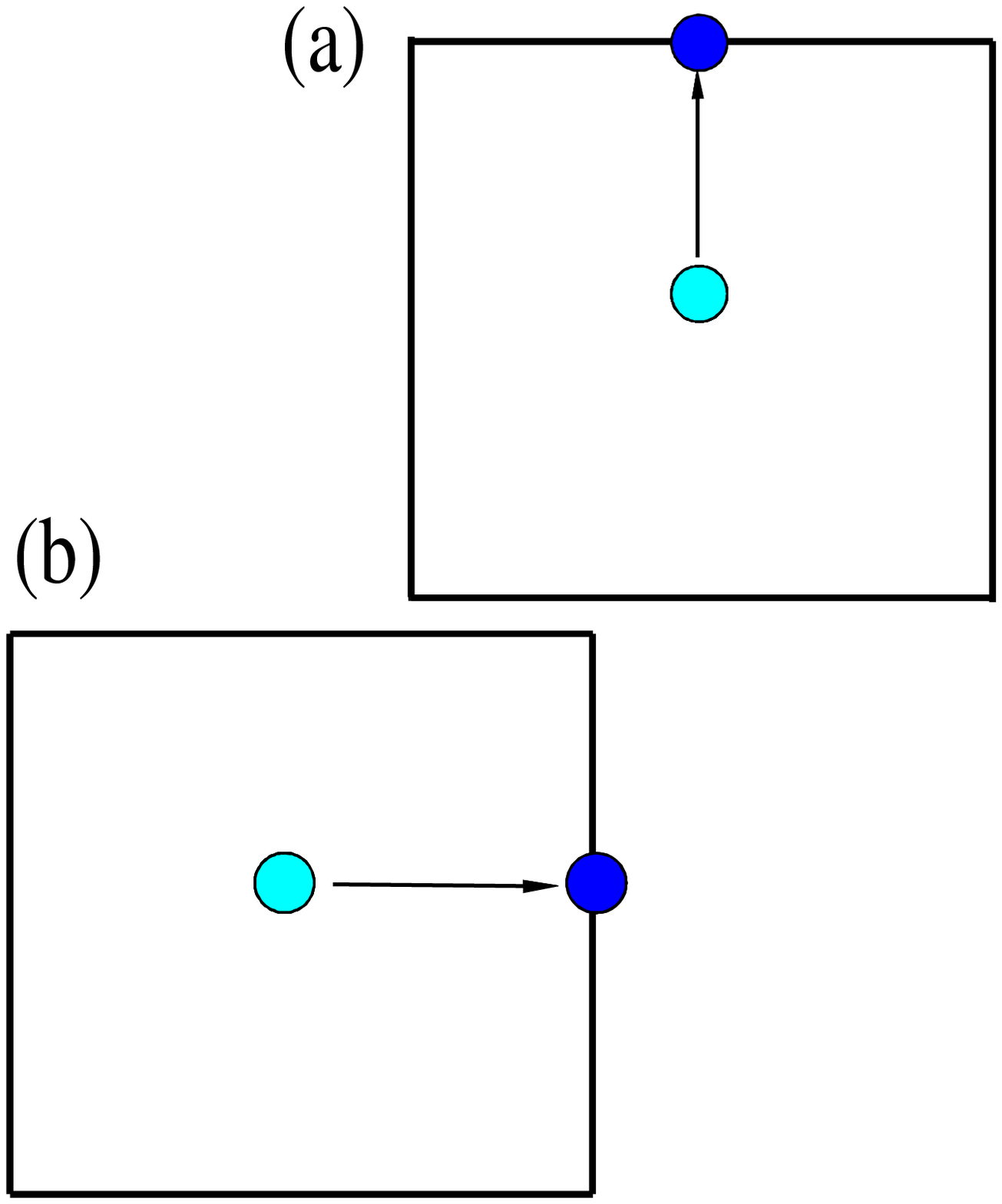}
\caption{\label{fig05}
}
\end{center}
Fundamental zone display for $DC4$ source.
Colors show two face points corresponding to one source
orientation with the angle $\Phi \ge 90^\circ$.
\end{figure}

\begin{figure}
\begin{center}
\includegraphics[width=0.75\textwidth]{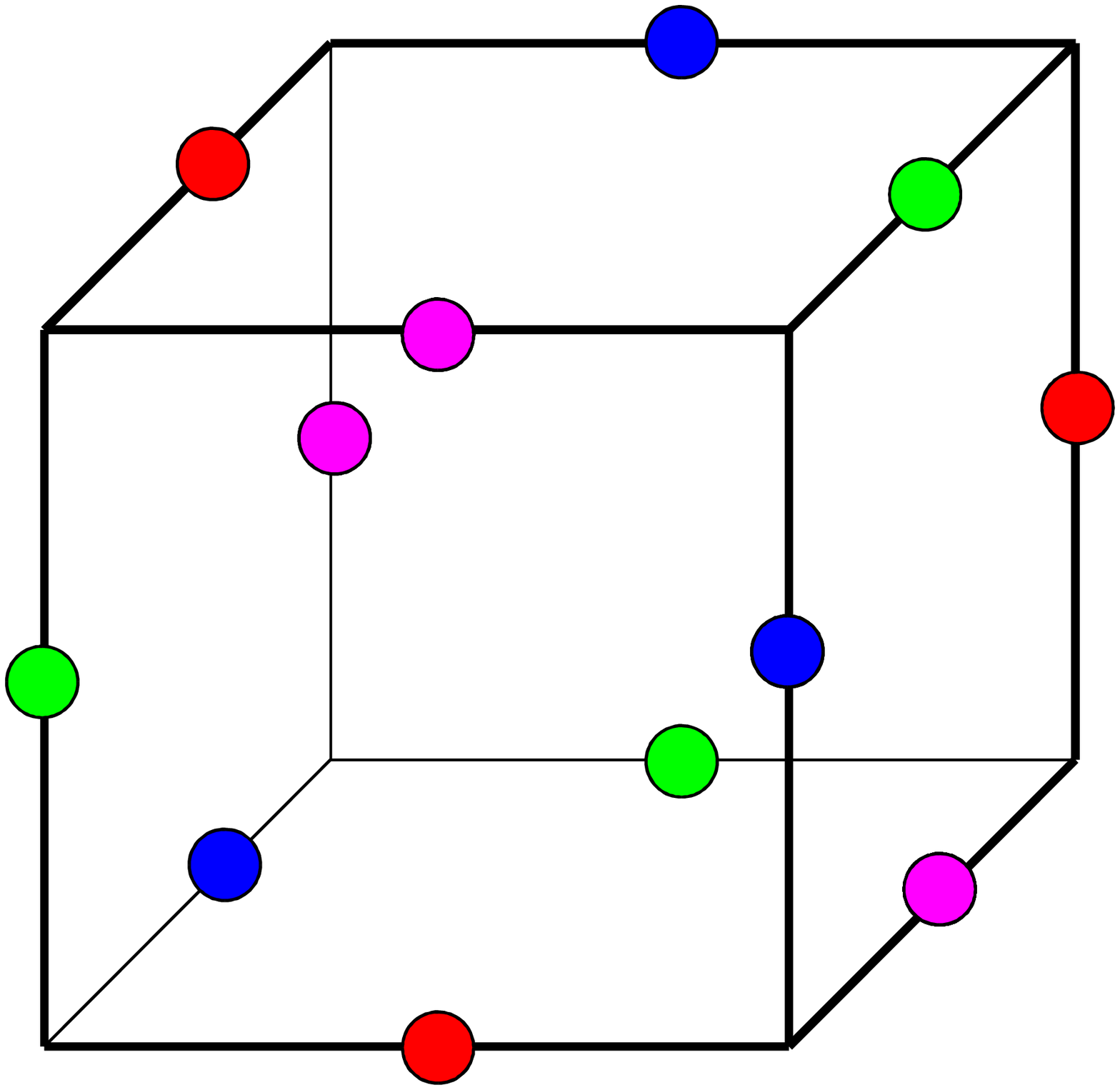}
\caption{\label{fig06}
}
\end{center}
Fundamental zone display for $DC4$ source.
Colors show four sets of three edge points corresponding to
one source disorientation with the angle $\Phi \approx
109.5^\circ$.
Compare to Fig.~\ref{fig05} where two edge points are result
of points moving on opposing faces.
The third point appears as it moves from outside the
fundamental cube to the third edge.
Three other sets of points are similarly produced.
\end{figure}

\begin{figure}
\begin{center}
\includegraphics[width=0.75\textwidth]{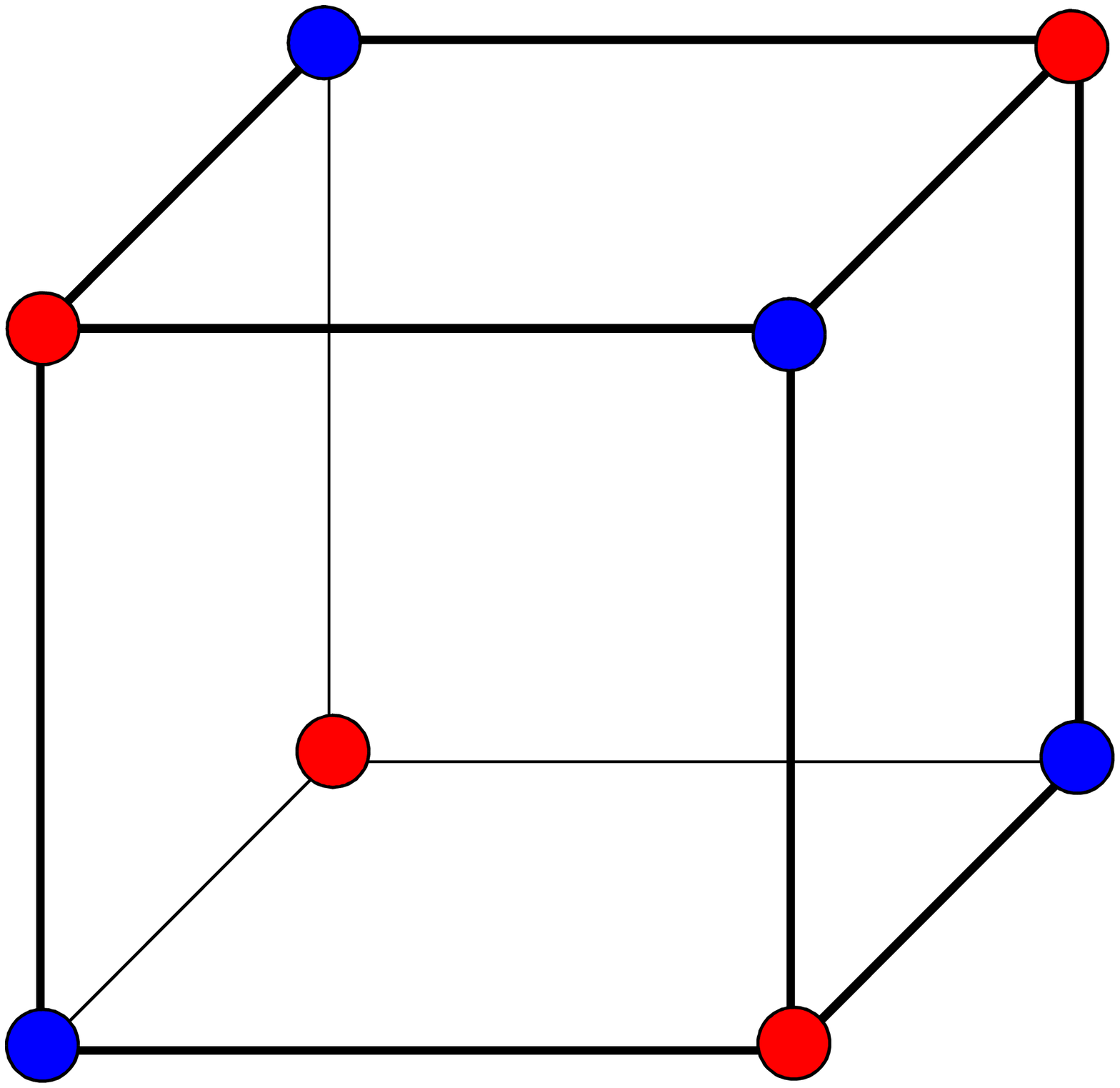}
\caption{\label{fig07}
}
\end{center}
Fundamental zone display for $DC4$ source.
Colors show two sets of four vertex points corresponding to
one source orientation with the angle $\Phi = 120^\circ$.
\end{figure}

\begin{figure}
\begin{center}
\includegraphics[width=0.75\textwidth]{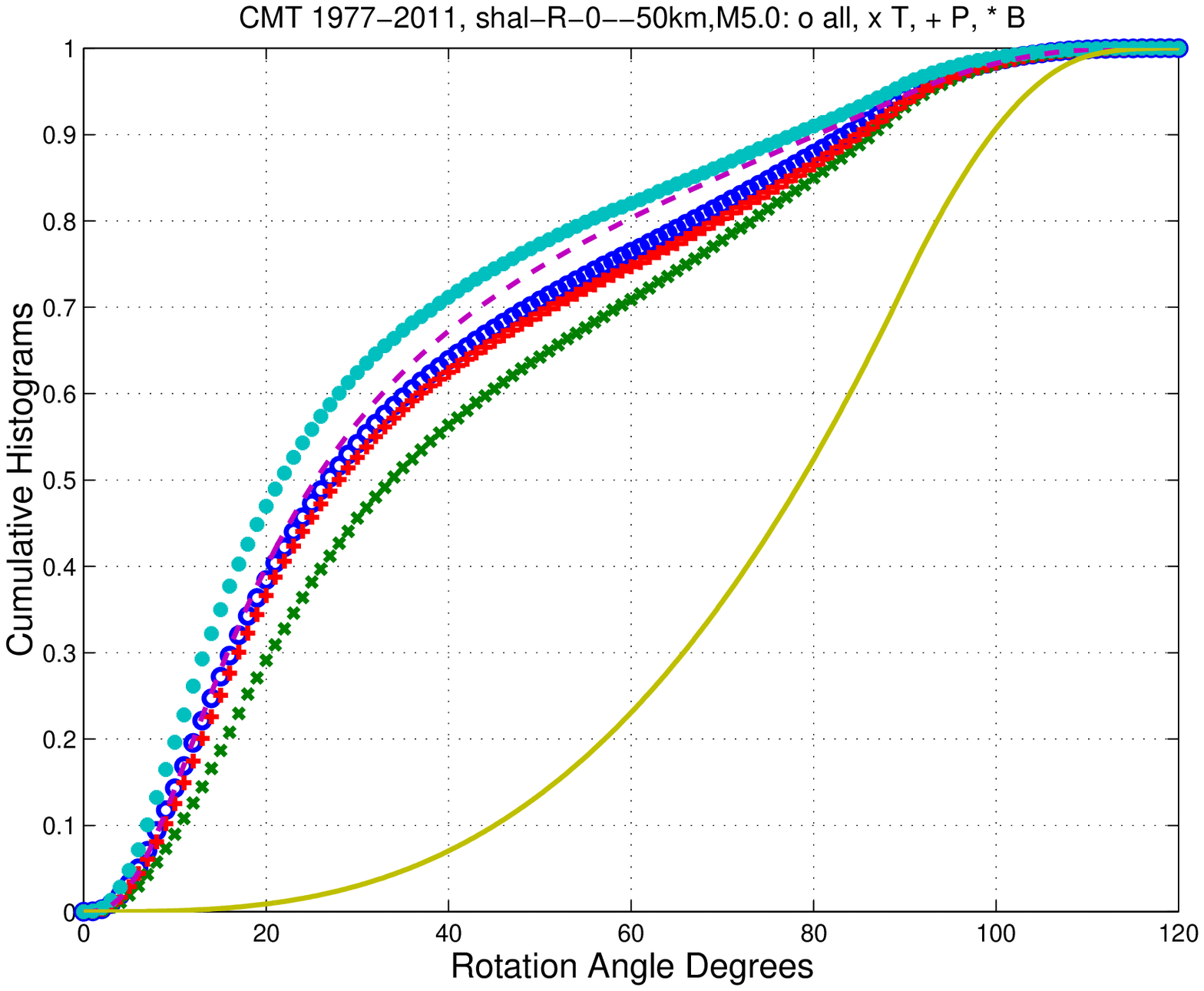}
\caption{\label{fig08}
}
\end{center}
Cumulative distributions of rotation angles for pairs of focal
mechanisms of shallow earthquakes (depth 0-70~km) in the GCMT
catalog 1977/01/01--2011/12/31; centroids are separated by
distances between 0-50~km, magnitude threshold $m_t = 5.0$.
The total number of events is 26,986.
Lines from left to right:
filled circles are centroids in 30$^\circ$ cones around the
{\bf b}-axis;
dashed line is for the Cauchy rotation with $\kappa =
0.1$;
circles -- all centroids;
crosses -- centroids in 30$^\circ$ cones around the {\bf
p}-axis;
x-signs -- centroids in 30$^\circ$ cones around the {\bf
t}-axis;
right solid line is for the random rotation.
\end{figure}

\begin{figure}
\begin{center}
\includegraphics[width=0.75\textwidth]{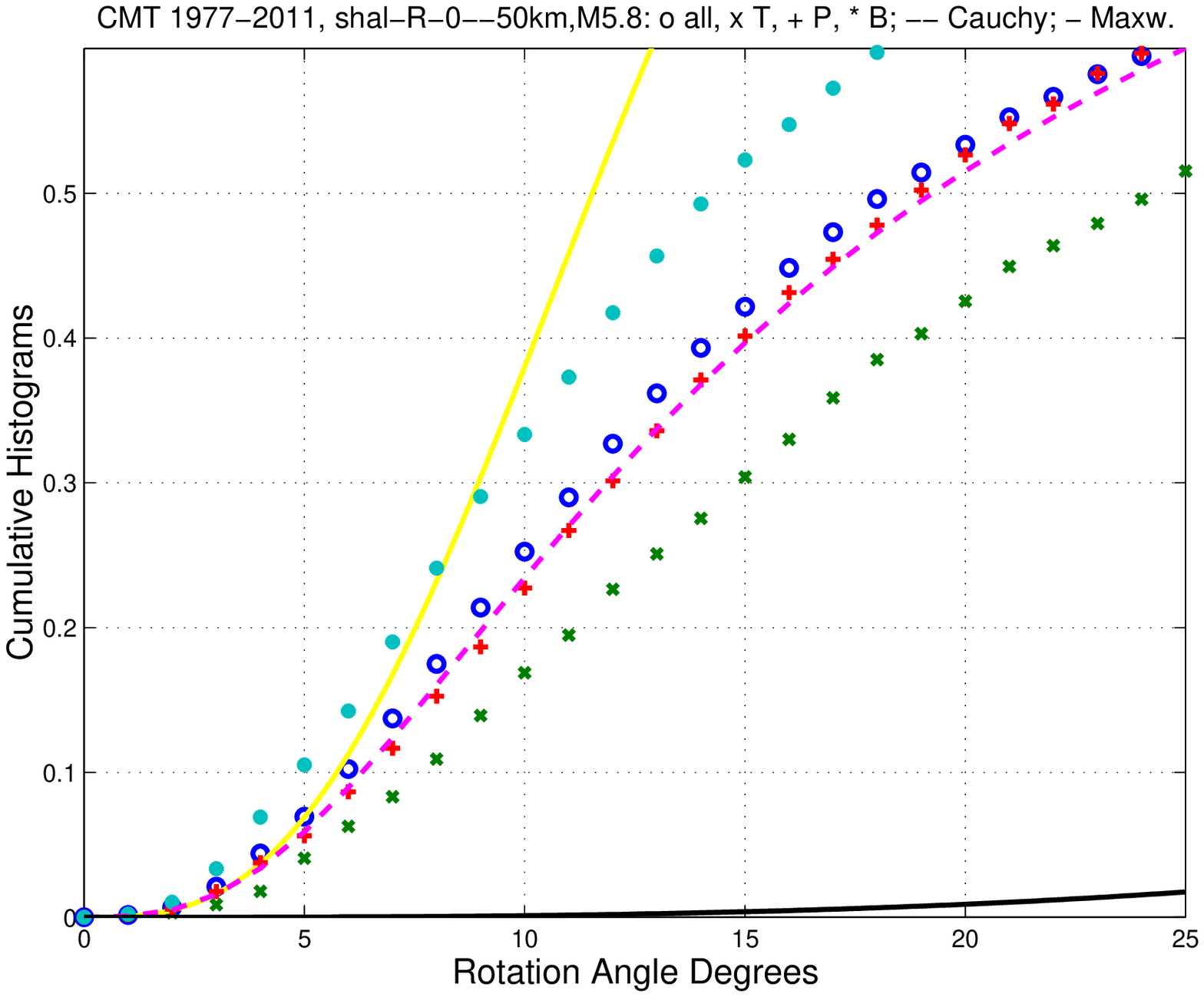}
\caption{\label{fig09}
}
\end{center}
Cumulative distributions of rotation angles for pairs of focal
mechanisms of shallow earthquakes (depth 0-70~km) in the GCMT
catalog 1977/01/01--2011/12/31; centroids are separated by
distances between 0-50~km, magnitude threshold $m_t = 5.8$.
The total number of events is 6,160.
Lines from left to right:
filled circles are centroids in 30$^\circ$ cones around the
{\bf b}-axis;
circles -- all centroids;
crosses -- centroids in 30$^\circ$ cones around the {\bf
p}-axis;
x-signs -- centroids in 30$^\circ$ cones around the {\bf
t}-axis;
dashed line is for the Cauchy rotation CDF with $\kappa =
0.075$;
left solid line is for the Maxwell rotation CDF (\ref{eq14})
with $\sigma_\Phi = 7.5^\circ$;
right solid line is for the random rotation.
\end{figure}

\begin{figure}
\begin{center}
\includegraphics[width=0.75\textwidth]{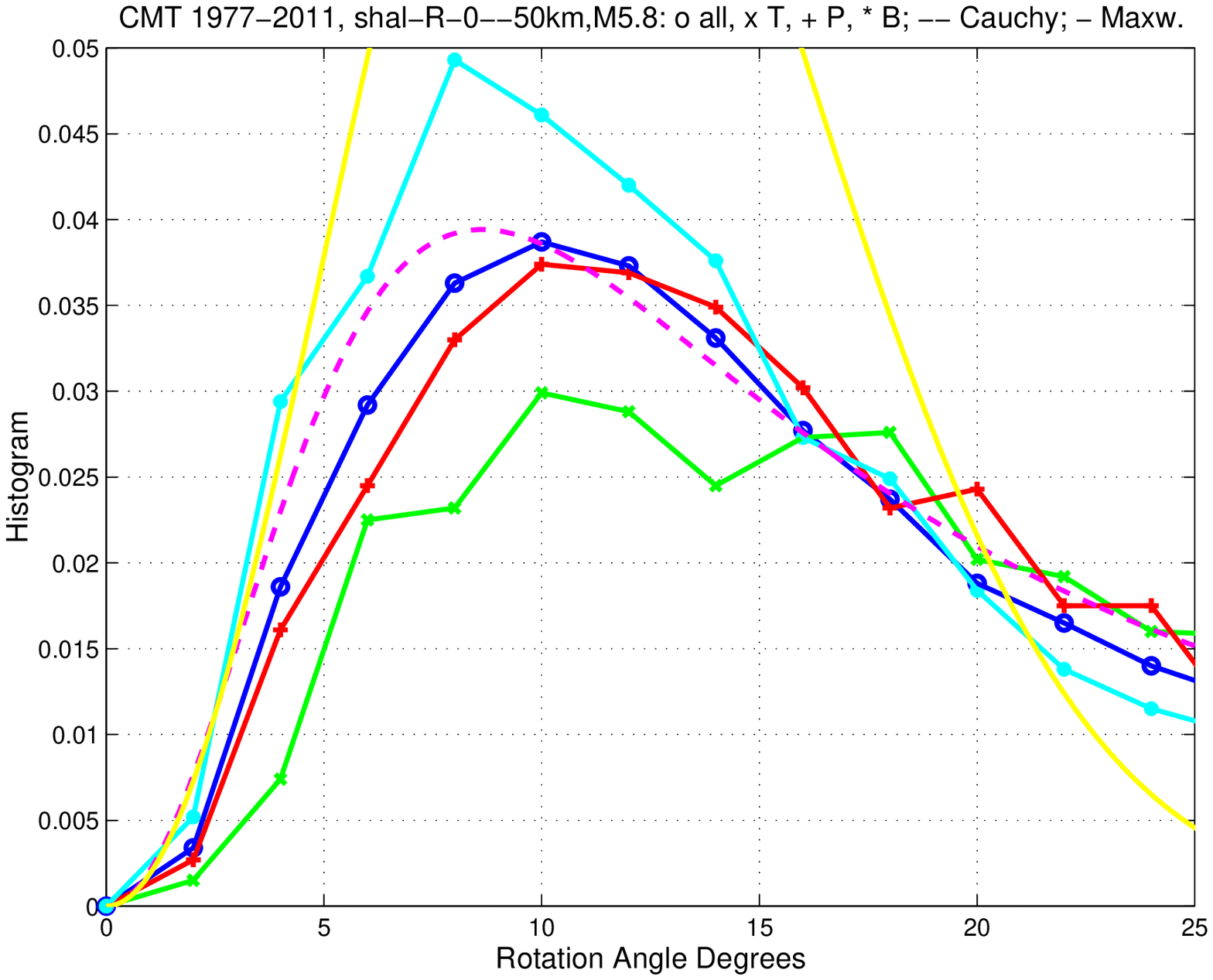}
\caption{\label{fig10}
}
\end{center}
Histograms of rotation angles for pairs of focal mechanisms
of shallow earthquakes (depth 0-70~km) in the GCMT catalog
1977/01/01--2011/12/31; centroids are separated by distances
between 0-50~km, magnitude threshold $m_t = 5.8$.
The total number of events is 6,160.
Lines from left to right:
filled circles are centroids in 30$^\circ$ cones around the
{\bf b}-axis;
circles -- all centroids;
crosses -- centroids in 30$^\circ$ cones around the {\bf
p}-axis;
x-signs -- centroids in 30$^\circ$ cones around the {\bf
t}-axis;
dashed line is for the Cauchy rotation PDF with $\kappa =
0.075$;
left solid line is for the Maxwell rotation PDF (\ref{eq13})
with $\sigma_\Phi = 7.5^\circ$.
\end{figure}

Mackenzie (1964) derived the distribution of the rotation axes
for cubic symmetry.
Morawiec (1996) obtained distributions of rotation axes for any
symmetric object encountered in material texture analysis.
Using his results we can write down the distribution of
rotation axes for the $D_2$ symmetry: the $DC4$ source.
The Mackenzie cell is shown in Fig.~\ref{fig04}.
We designate the coordinate axes as $x_i, \, i=1,2,3$, and
the distribution depends on distance from the origin.
As seen in Fig.~\ref{fig04}, the distribution should have a
3-fold cyclic symmetry $C_3$ around the origin or around the
cube vertex.
Then the PDF for the axes density is
\be
p \, (\rho) \ = \ { {16} \over {\pi^2} }
\left [ \, \arctan (\rho) - \rho \, /(1 + \rho^2) \right ] \, ,
\label{eq15}
\ee
where the distance $\rho = \sqrt {x_1^2 + x_2^2 + x_3^2 } $.
For small value of $\rho$, $ p \, (\rho) \ \propto \ \rho^3$.

\begin{figure}
\begin{center}
\includegraphics[width=0.75\textwidth]{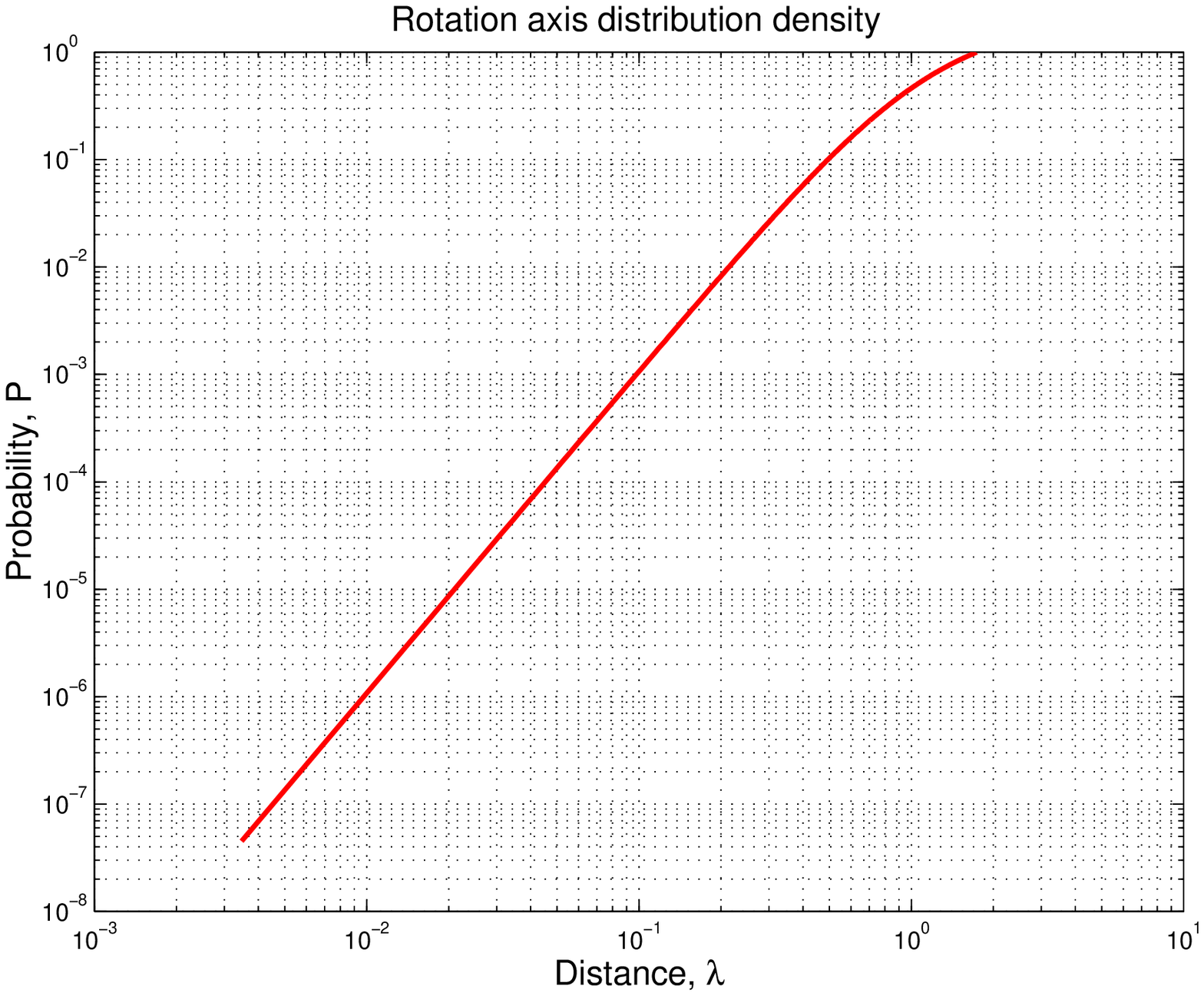}
\caption{\label{fig11}
}
\end{center}
Probability density function for rotation axes distribution.
\end{figure}

Fig.~\ref{fig11} displays the $ p \, (\rho) $ density.
Its values at $\rho=1$, $\rho=\sqrt 2$, and $\rho=\sqrt 3$
correspond to appropriate values for the $D_2$ symmetry
(Morawiec 1996, Table~2).
These $\rho$-values correspond to the rotation angles $\Phi =
90^\circ, 109.5^\circ$, and $120.0^\circ$, respectively.
For $\Phi \le 90^\circ$ the rotation axes are distributed
uniformly over the $S^2$ sphere, but they intersect the sphere
near the cube vertex close to $\Phi = 120^\circ$.

\begin{figure}
\begin{center}
\includegraphics[width=0.75\textwidth]{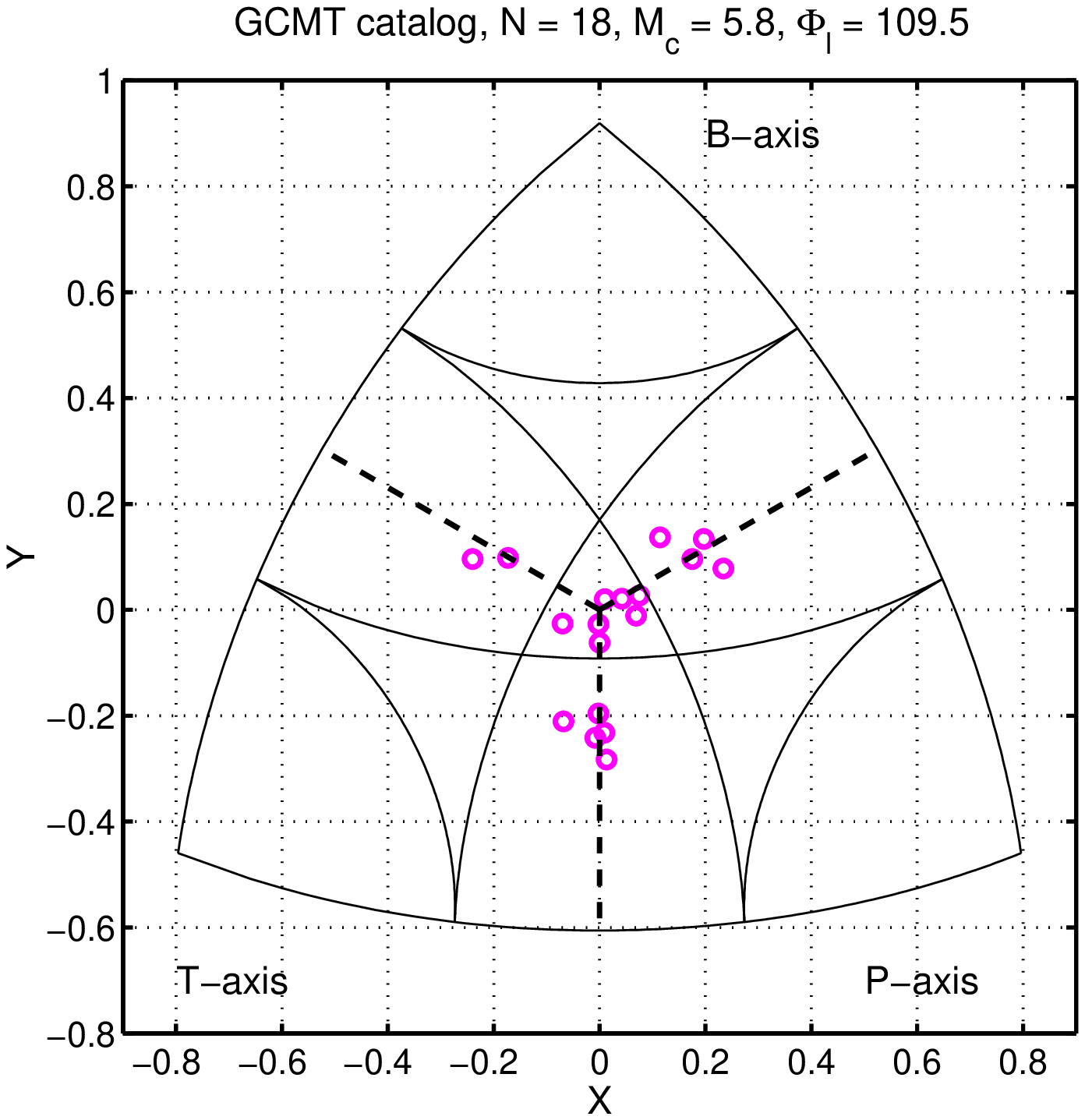}
\caption{\label{fig12}
}
\end{center}
Distributions of rotation poles for pairs of focal mechanisms
of shallow earthquakes in the GCMT catalog.
Centroids are separated by distances between 0-50~km;
magnitude threshold $m_t = 5.8$;
the rotation angle $109.5^\circ \le \Phi \le 120^\circ $.
\end{figure}

In Fig.~\ref{fig12} we show the distribution of the rotation
poles for the second earthquake focal mechanism on a reference
sphere of the first event.
Because of the symmetry of the $DC4$ source, we reflect the
point pattern on our reference sphere at the planes
perpendicular to all axes.
Thus, the distribution can be shown on a spherical octant.
We use the Lambert azimuthal equal-area projection.
The points concentrate near the projections of the far edges
of the MacKenzie cell (see Fig.~\ref{fig04}) and around the
cube vertex which corresponds to the disorientation angle
$\Phi=120^\circ$.

\subsection{Rodrigues space statistics and display }
\label{stat3}
A major problem in the orientation visualizing is the high
dimensionality of the 3-D rotation space: the orthogonal
matrices are characterized by nine values, the seismic
moment tensor requires five or six variables, and the
normalized quaternion needs four values.
The real number of degrees of freedom for a 3-D rotation is
three.
Thus, in principle, an orientation distribution can be
shown in a 3-D diagram.

Frank (1988), Neumann (1992) and Morawiec \& Field (1996)
propose using the Rodrigues vector space to display the
disorientation of symmetric objects in a fundamental zone as a
point in the space.
The point coordinates are calculated as follows: the length of
a vector $\zeta$ is
\be
\zeta \ = \ \tan (\Phi/2) \, .
\label{eq16}
\ee
where $\Phi_{\rm min}$ is the rotation angle.
Three coordinates of a point in the zone are
\ba
x_1 \ = \ & \zeta \times \sin(\theta) \, \sin(\phi);
\nonumber\\
x_2 \ = \ & \zeta \times \sin(\theta) \, \cos(\phi);
\nonumber\\
x_3 \ = \ & \zeta \times \cos(\theta) \, ,
\label{eq17}
\ea
where $\theta$ is the colatitude, and $\phi$ is the azimuth of
the rotation axis.
As shown in Fig.~\ref{fig04}, we identify $x_1$ with the {\bf
p}-axis; similarly $x_2 = {\bf t}$ and $x_3 = {\bf b}$.

We obtain a distribution diagram for a set of disorientations.
One way to display such a diagram of the fundamental zone is
through stereo-pairs (Neumann 1992).
Morawiec \& Field (1996) display the distribution of
misorientation parameters by points in some parallel sections
of the fundamental zone.

\begin{figure}
\begin{center}
\includegraphics[width=0.75\textwidth]{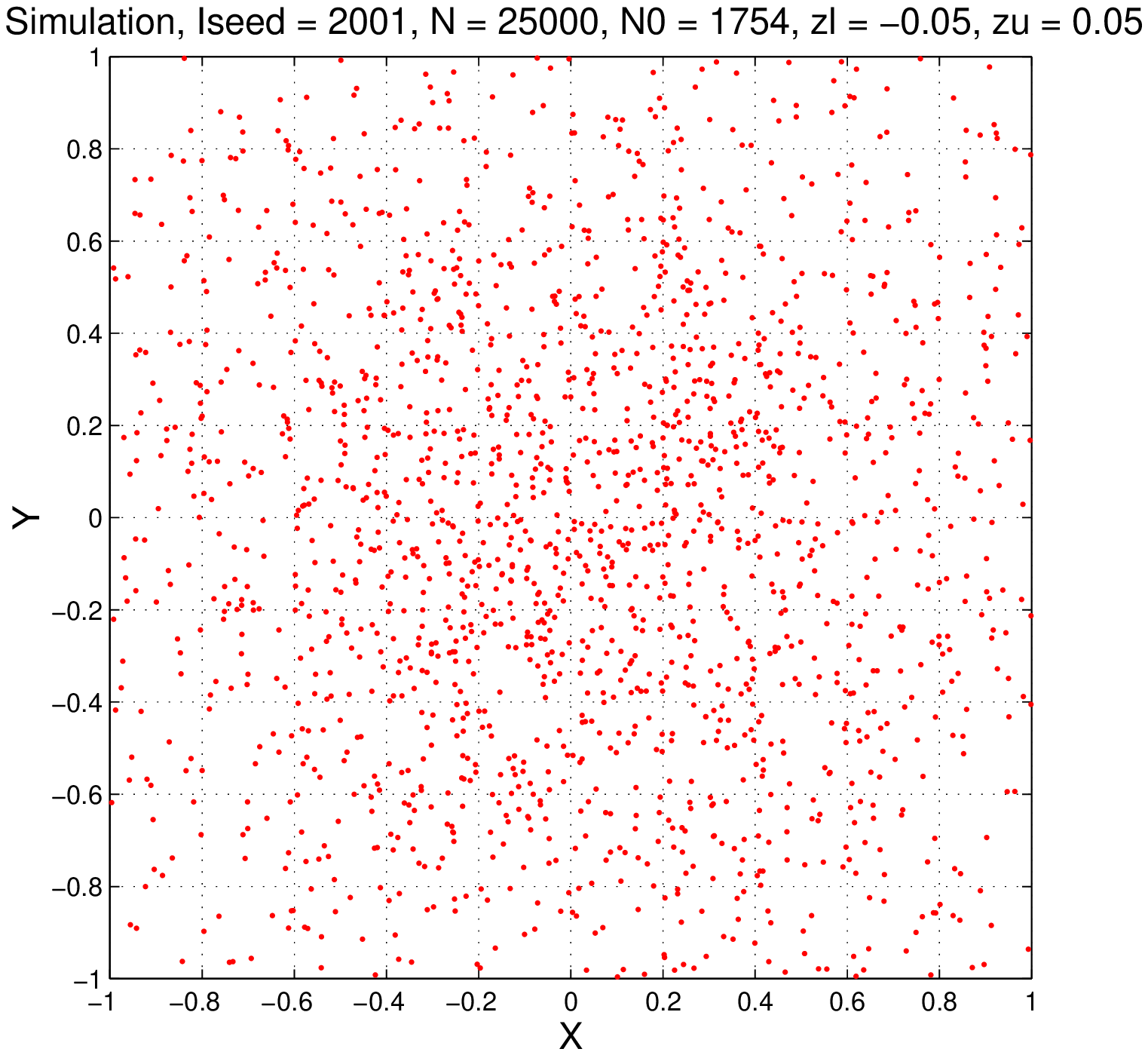}
\caption{\label{fig13}
}
\end{center}
Orientation distribution in the fundamental zone of the
Rodrigues space for randomly rotated $DC4$ sources.
The points are shown in the central section of the fundamental
zone $0.05 \ge x_3 \ge -0.05$ (see Fig.~\ref{fig04}).
The total point number is 25,000; 1,754 points are in the
central section.
\end{figure}

\begin{figure}
\begin{center}
\includegraphics[width=0.75\textwidth]{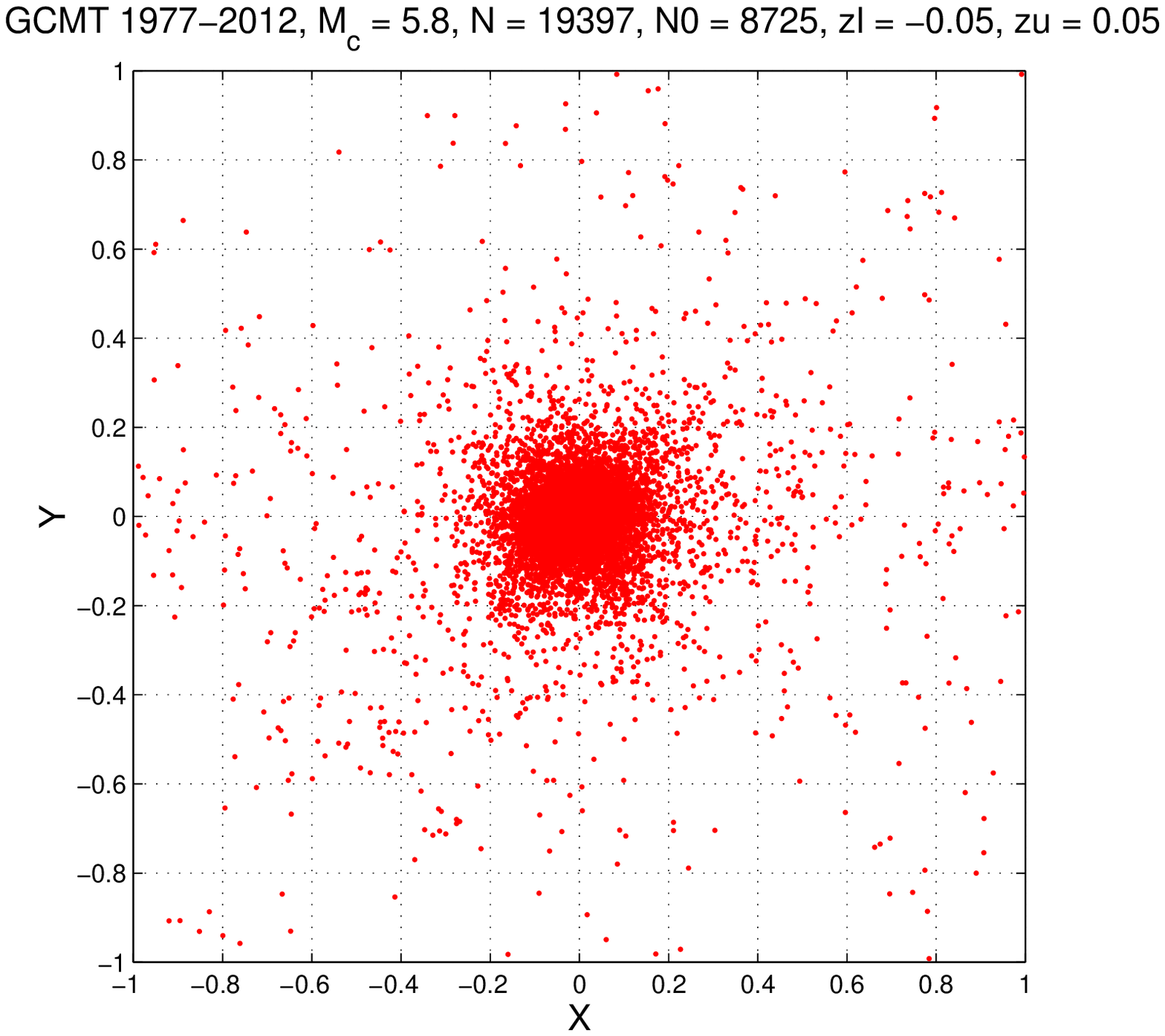}
\caption{\label{fig14}
}
\end{center}
Orientation distribution in the fundamental zone of the
Rodrigues space for shallow earthquakes in the GCMT
catalog.
Centroids are separated by distances between 0-50~km;
magnitude threshold $m_t = 5.8$;
the total number of events is 6,160.
The points are shown in the central section of the fundamental
zone $0.05 \ge x_3 \ge -0.05$ (see Fig.~\ref{fig04}).
The total number of earthquake pairs is 19,397; 8,725 pair
points are in the central section.
\end{figure}

Fig.~\ref{fig13} shows a distribution for randomly rotated
$DC4$ sources in a central section of the fundamental zone of
the Rodrigues space.
Fig.~\ref{fig14} displays a similar distribution of the
earthquake focal mechanism orientation in the GCMT catalog.
As Figs.~\ref{fig08}-\ref{fig10} demonstrate, the distribution
of the rotation angles for earthquake sources is strongly
concentrated close to $0^\circ$.
If we compare Fig.~\ref{fig13} and Fig.~\ref{fig14}, this
concentration is marked in the fundamental zone display.

Table~\ref{Table1} summarizes earthquake focal mechanism
disorientation patterns in the fundamental zone of $DC4$.
The total number of events $N$ with the magnitude above the
threshold is shown, as well as the total number of pairs
$N_p$ with centroids separated by less than 50~km.
$N_c$ is the pair numbers in the central zone shown in
Fig.~\ref{fig14}.
Although the central section occupies only 5\% of the zone,
close to 50\% of the pairs are there due to a tight
concentration of rotation angles near the zero value.
For simulated events in the central zone, the number $N_c$ is
about 7\% of the total (see Fig.~\ref{fig13}).

We also display the correlation coefficients of the point
scatter field.
Whereas for earthquake focal mechanisms the coefficients
$\rho_{bp}$ and $\rho_{bt}$ are close to zero, the
$\rho_{pt}$ and $\rho_{pt}^\prime$ coefficients are non-zero,
testifying to a certain pattern of focal mechanism rotation.
Fig.~11 in Kagan (2009) also shows that rotation axes are
concentrated closer to the {\bf t}-axis.
All correlation coefficients for the simulated mechanisms are
around zero.
A more detailed statistical analysis of this pattern
will be carried out in our future work.

The values of the average rotation angle and its standard
deviation ($ \, \overline \Phi \pm \sigma_\Phi$) show that for
larger earthquakes both variables are smaller.
This may be caused by a higher accuracy in determining focal
mechanism for stronger shocks (Kagan 2003).
For simulated focal mechanisms, the $\, \overline \Phi \pm
\sigma_\Phi$ values are close to the theoretical estimates
for orthorhombic symmetry (Grimmer 1979).

\section{Discussion }
\label{disc}
Quantitative study of earthquake focal mechanisms is an
important prerequisite for understanding earthquake rupture.
Though these investigations began in the mid 1950s,
publications have been mostly descriptive until now;
relatively little modelling and rigorous statistical analysis
have been performed.
A major difficulty in analyzing focal mechanisms is both the
high dimensionality and non-commutativity of the 3-D
rotations.
This presents a major challenge in analyzing a set of focal
mechanisms.

Several papers (Kagan, 1992, 2000, 2009) have investigated
statistical distributions of earthquake focal mechanisms.
We found that the disorientation angle is close to zero for
spatially close earthquakes, and the angle decreases if
the inter-earthquake time interval approaches zero.
We also showed (Kagan, 2009) that the CLVD component of focal
mechanism tensor is either zero or close to zero for most
geometric barriers proposed as common features in an
earthquake fault system.

However, the major challenges in describing and understanding
the distributions of focal mechanisms still remain.
As we see from Figs.~\ref{fig08}-\ref{fig10}, the angle
distribution is not axially symmetric: in certain directions
$\Phi_{\rm min}$ is larger than in others.
Thus, the distributions used to approximate the angle
pattern, like the rotational Cauchy distribution
(Eqs.~\ref{eq8}-\ref{eq9}), need to be made more complex.

The distribution of rotation axes was not investigated as
thoroughly as that for disorientation angles.
There is still no theoretical model for approximating
empirical data, but applying the Rodrigues space may render
such analysis more manageable.

However, even these limited results contribute significantly
to understanding of earthquake focal mechanism properties and
allows certain quantitative applications for seismic risk
evaluation.
Kagan \& Jackson (2011) explain that the forecasted tensor
focal mechanism enables calculating an ensemble of
seismograms for each point of interest on the Earth's surface.
Moreover, the focal mechanism distribution allows us to
estimate fault plane orientation for past earthquakes, through
which we can identify a preferred rupture direction for future
events.

The angle $\Phi_{\rm min}$ has also been used to directly
compare moment tensors from two different earthquakes
(Okal {\it et al.}\ 2011).
It has as well been applied in comparing moment tensors
computed for the same events through different techniques
(Frohlich \& Davis 1999; Pondrelli {\it et al.}\ 2007; Yang
{\it et al.}\ 2012).
Such comparisons can help refine the moment tensor algorithms
and lower their computational cost, since they can reveal the
relative importance of various assumptions implicit in the
algorithms.

\section{Conclusions }
\label{conc}

{\phantom {\bigtensl {WEB Home Pages: }}}$\,$
\vskip - 1.5 cm

$\bullet$
1.
The symmetry properties of an earthquake double-couple focal
mechanism are considered.
Given available seismological and geologic information,
three symmetries are possible: $D_2$ (dihedral symmetry),
$C_2$ (cyclic symmetry), and $I$ (identity).
Determining the orientation or disorientation of the source
depends on its symmetry.

$\bullet$
2.
Quaternion representation is the most convenient tool for
analysing a double-couple 3-D rotation.
A 3-D rotation requires at least three degrees of freedom
for its characterization.

$\bullet$
3. Several theoretical distributions to describe a 3-D
rotation of double-couples are presented: random rotation,
rotational Cauchy, and von Mises-Fisher.

$\bullet$ 4. The Rodrigues space, so extensively used in material
texture analysis is applied to display and analyze earthquake focal
mechanism distribution. This space allows us to represent 3-D
patterns of how symmetric objects are oriented.

$\bullet$
5. We illustrate the proposed methods by statistically
analyzing the GCMT catalog of earthquake focal mechanisms.

\subsection* {Acknowledgments
} \label{Ackn} I am grateful to Dave Jackson, Paul Davis, and Peter
Bird of UCLA, as well as Maximilian Werner and Men Meier of ETH
Zurich for useful discussion and suggestions. I thank Kathleen
Jackson for editing and significant improvements in the text. The
author appreciates partial support from the National Science
Foundation through grants EAR-0944218, and EAR-1045876, as well as
from the Southern California Earthquake Center (SCEC). SCEC is
funded by NSF Cooperative Agreement EAR-0529922 and USGS Cooperative
Agreement 07HQAG0008. Publication 0000, SCEC.

\pagebreak

\centerline { {\sc References} }
\vskip 0.1in
\parskip 1pt
\parindent=1mm
\def\reference{\hangindent=22pt\hangafter=1}

\reference
Aki, K.~\& Richards, P.\ 2002.
{\sl Quantitative Seismology},
2nd ed., Sausalito, Calif., University Science Books, 700~pp.

\reference
Altmann, S.\ L.\ 1986.
{\sl Rotations, Quaternions and Double Groups},
Clarendon Press, Oxford, pp.\ 317.

\reference
Chen, P., T.H. Jordan, Li Zhao 2010.
Resolving fault plane ambiguity for small earthquakes,
{\sl Geophys.\ J. Int.}, {\bf 181}(1), 493-501, DOI:
10.1111/j.1365-246X.2010.04515.x.

\reference
Dunn, F. \& I. Parberry 2011.
{\sl 3D Math Primer for Graphics and Game Development},
A.K. Peters/CRC Press, 846~pp.

\reference
Ekstr\"om, G., A. M. Dziewonski, N. N. Maternovskaya
\& M. Nettles 2005.
Global seismicity of 2003: Centroid-moment-tensor solutions
for 1087 earthquakes,
{\sl Phys.\ Earth planet.\ Inter.}, {\bf 148}(2-4), 327-351.

\reference
Frank, F. C.\ 1988.
Orientation mapping,
{\sl Metall.\ Trans.\ A}, {\bf 19}, 403-408.

\reference
Frohlich, C. \& S.\ D.\ Davis 1999.
How well constrained are well-constrained T, B, and P axes in
moment tensor catalogs?,
{\sl J.\ Geophys.\ Res.}, {\bf 104}, 4901-4910.

\reference
Gasperini, P. \& G. Vannucci 2003.
FPSPACK: a package of FORTRAN subroutines to manage earthquake
focal mechanism data,
{\sl Computers \& Geosciences}, {\bf 29}(7), 893-901, DOI:
10.1016/S0098-3004(03)00096-7.

\reference
Grimmer, H.\ 1979.
Distribution of disorientation angles if all relative
orientations of neighboring grains are equally probable,
{\sl Scripta Metallurgica}, {\bf 13}(2), 161-164,
DOI: 10.1016/0036-9748(79)90058-9.

\reference
Handscomb, D.\ C.\ 1958.
On the random disorientation of two cubes,
{\sl Can.\ J.\ Math.}, {\bf 10}, 85-88.

\reference
Hanson, A. J.\ 2006.
{\sl Visualizing Quaternions},
San Francisco, Calif., Elsevier, pp.~498.

\reference
Heinz, A. \& Neumann, P.\ 1991.
Representation of orientation and disorientation data for
cubic, hexagonal, tetragonal and orthorhombic crystals,
{\sl Acta Cryst.\ A}, {\bf 47}, 780-789.

\reference
Horn, B. K. P.\ 1987.
Closed-form solution of absolute orientation using unit
quaternions,
{\sl J. Opt.\ Soc.\ Am.\ A}, {\bf 4}(4), 629-642.

\reference
Kagan, Y.~Y.\ 1982.
Stochastic model of earthquake fault geometry,
{\sl Geophys.\ J.\ Roy.\ astr.\ Soc.}, {\bf 71}(3), 659-691.

\reference
Kagan, Y.~Y.\ 1990.
Random stress and earthquake statistics: Spatial dependence,
{\sl Geophys.\ J. Int.}, {\bf 102}(3), 573-583.

\reference
Kagan, Y.~Y.\ 1991.
3-D rotation of double-couple earthquake sources,
{\sl Geophys.\ J. Int.}, {\bf 106}(3), 709-716.

\reference
Kagan, Y.~Y.\ 1992.
Correlations of earthquake focal mechanisms,
{\sl Geophys.\ J. Int.}, {\bf 110}(2), 305-320.

\reference
Kagan, Y. Y.\ 2000.
Temporal correlations of earthquake focal mechanisms,
{\sl Geophys.\ J. Int.}, {\bf 143}(3), 881-897.

\reference
Kagan, Y. Y., 2003.
Accuracy of modern global earthquake catalogs,
{\sl Phys.\ Earth Planet.\ Inter.}, {\bf 135}(2-3),
173-209, doi:10.1016/S0031-9201(02)00214-5.

\reference
Kagan, Y. Y.\ 2007.
Simplified algorithms for calculating double-couple rotation,
{\sl Geophys.\ J. Int.}, {\bf 171}(1), 411-418,
doi: 10.1111/j.1365-246X.2007.03538.x.

\reference
Kagan, Y. Y.\ 2009.
On the geometric complexity of earthquake focal zone and fault
systems: A statistical study,
{\sl Phys.\ Earth Planet.\ Inter.}, {\bf 173}(3-4), 254-268,
doi: 10.1016/j.pepi.2009.01.006.

\reference
Kagan, Y. Y. \& Jackson, D. D.\ 2011.
Global earthquake forecasts,
{\sl Geophys.\ J. Int.}, {\bf 184}(2), 759-776,
doi: 10.1111/j.1365-246X.2010.04857.x

\reference
Krieger, L. \& S. Heimann 2012.
MoPaD -- Moment tensor plotting and decomposition: A tool for
graphical and numerical analysis of seismic moment tensors,
{\sl Seismol.\ Res.\ Lett.}, {\bf 83}(3), 589-595,
doi:10.1785/gssrl.83.3.589.

\reference
Kuipers, J. B.\ 1999.
{\sl Quaternions and Rotation Sequences: A Primer
with Applications to Orbits, Aerospace and Virtual Reality},
Princeton, Princeton Univ.\ Press., 400~pp.

\reference
Mackenzie, J.\ K.\ 1958.
Second paper on statistics associated with the random
disorientation of cubes,
{\sl Biometrika}, {\bf 45}, 229-240.

\reference
Mackenzie, J.K.\ 1964.
Distribution of rotation axes in random aggregate of cubic
crystals,
{\sl Acta Metallurgica}, {\bf 12}(2), 223-225, DOI:
10.1016/0001-6160(64)90191-9.

\reference
Mardia, K.\ V. \& P. E. Jupp 2000.
{\sl Directional Statistics},
Chichester, New York, Wiley, 429~pp.

\reference
Mason, J.K. \& C.A. Schuh 2009.
The generalized Mackenzie distribution: Disorientation
angle distributions for arbitrary textures,
{\sl Acta Materialia}, {\bf 57}, 4186-4197.

\reference
McGuire, J. J., Li Zhao \& Jordan, T. H.\ 2001.
Teleseismic inversion for the second-degree moments of
earthquake space-time distributions,
{\sl Geophys.\ J. Int.}, {\bf 145}, 661-678.

\reference
Meister, L. \& H. Schaeben 2005.
A concise quaternion geometry of rotations,
{\sl Math.\ Meth.\ Appl.\ Sci.}, {\bf 28}, 101-126.

\reference
Moakher, M.\ 2002.
Means and averaging in the group of rotations,
{\sl SIAM J. Matrix Anal. Appl.}, {\bf 24}(1), 1-16.

\reference
Morawiec, A.\ 1995.
Misorientation-angle distribution of randomly oriented
symmetrical objects,
{\sl J. Applied Crystallography}, {\bf 28}(3), 289-293, DOI:
10.1107/S0021889894011088.

\reference
Morawiec, A., 1996.
Distributions of rotation axes for randomly oriented symmetric
objects,
{\sl J. Applied Crystallography}, {\bf 29}(2), 164-169, DOI:
10.1107/S0021889895013641.

\reference
Morawiec, A.\ 2004.
{\sl Orientations and Rotations: Computations in
Crystallographic Textures},
Springer/Berlin, New York, pp.~200.

\reference
Morawiec, A. \& Field, D. P.\ 1996.
Rodrigues parameterization for orientation
and misorientation distributions,
{\sl Philos.\ Mag.\ A}, {\bf 73}(4), 1113-1130.

\reference
Neumann, P., 1992.
The role of geodesic and stereographic projections
for the visualization of directions, rotations, and textures,
{\sl Phys. Stat. Sol. (a)}, {\bf 131}, 555-567.

\reference
Okal, E. A., Borrero, J. C., \& Chagui-Goff, C. 2011.
Tsunamigenic predecessors to the 2009 Samoa earthquake,
{\sl Earth-Sci. Rev.}, {\bf 107}, 128-140.

\reference
Pondrelli S., Salimbeni, S., Morelli, A., Ekstr\"om, G.,
Boschi, E. 2007.
European-Mediterranean regional centroid moment tensor
catalog: Solutions for years 2003 and 2004,
{\sl Phys.\ Earth Planet.\ Inter.}, {\bf 164}(1-2), 90-112.

\reference
Schaeben, H.\ 1996.
Texture approximation or texture modelling with components
represented by the von Mises-Fisher matrix distribution on
$SO(3)$ and the Bingham distribution on $S^4_+$,
{\sl J. Appl. Cryst.}, {\bf 29}, 516-525.

\reference
Schaeben, H.\ 2010.
Special issue on spherical mathematics and statistics,
{\sl Math Geosci.}, {\bf 42}, 727-730,
DOI 10.1007/s11004-010-9304-7.

\reference
Shepperd, S. W.\ 1978.
Quaternion from rotation matrix,
{\sl J. Guidance and Control}, {\bf 1}, 223-224.

\reference
Snoke, J. A.\ 2003.
FOCMEC: FOcal MEChanism determinations,
{\sl International Handbook of Earthquake and Engineering
Seismology} (W. H. K. Lee, H. Kanamori, P. C. Jennings \& C.
Kisslinger, Eds.), Academic Press, San Diego, Chapter 85.12.
pp.~1629-1630.

\reference
Yang, W., E. Hauksson, \& P. M. Shearer 2012.
Computing a large refined catalog of focal mechanisms for
southern California (1981-2010): Temporal stability of the
style of faulting,
{\sl Bull.\ Seismol.\ Soc.\ Amer.}, {\bf 102}, 1179-1194,
doi:10.1785/0120110311.

\reference
Zolotarev, V.\ M.\ 1986.
{\sl One-Dimensional Stable Distributions},
Amer.\ Math.\ Soc., Providence, R.I., pp.~284;
Russian original 1983.

\clearpage

\newpage

\renewcommand{\baselinestretch}{1.25}

\begin{table}
\caption{Properties of disorientation point scatter in the
fundamental zone of a $DC4$ source } \vspace{10pt} \label{Table1}
\begin{tabular}{lrrrrrrrrr}
\hline
& & & & & & & & & \\[-15pt]
\multicolumn{1}{c}{\#}&
\multicolumn{1}{c}{$m_t$}&
\multicolumn{1}{c}{$N$}&
\multicolumn{1}{c}{$N_p$}&
\multicolumn{1}{c}{$N_c$}&
\multicolumn{1}{c}{$\rho_{pt}$}&
\multicolumn{1}{c}{$\rho_{bp}$}&
\multicolumn{1}{c}{$\rho_{bt}$}&
\multicolumn{1}{c}{$\rho_{pt}^\prime$}&
\multicolumn{1}{c}{$\, \overline \Phi \pm \sigma_{\Phi} $ }
\\[2pt]
\hline
& & & & & & & & & \\[-15pt]
1 & 5.6 & 9,615 & 43,611 & 18,381 & 0.126 & 0.026 & 0.013 & 0.170 & 31.6 $\pm$ 27.5 \\
2 & 5.8 & 6,160 & 19,367 & 8,725 & 0.117 & 0.022 & $-$0.001 & 0.127 & 29.9 $\pm$ 26.9 \\
3 & 6.25 & 2,154 & 2,741 & 1,244 & 0.274 & $-$0.016 & $-$0.026 & 0.257 & 28.5 $\pm$ 26.1 \\
4 & Simul. & 25,000 & 25,000 & 1,754 & $-$0.012 & $-$0.002 & 0.002 & $-$0.015 & 75.0 $\pm$ 21.0 \\
\hline
\end{tabular}


\bigskip
{\sl Notes:}
The GCMT catalog time interval is 1977/1/1--2011/12/31.
$N$ is the total number of events with magnitude $m \ge m_t$;
$N_p$ is the total number of event pairs;
$N_c$ is the number of event pairs in the central section;
$\rho_{pt}$, $\rho_{bp}$, and $\rho_{bt}$ are
the correlation coefficients for all points;
$\rho_{pt}^\prime$ is the correlation coefficient for all
points within the central section;
$\, \overline \Phi \pm \sigma_{\Phi} $ is the average
disorientation angle and its standard deviation.
\hfil\break
\vspace{5pt}
\end{table}

\newpage

\clearpage

\renewcommand{\baselinestretch}{1.25}

\renewcommand{\baselinestretch}{1.75}

\parindent=0mm

\newpage

\end{document}